\documentclass[letterpaper,12pt,notoc]{JHEP3}
\def\pd{\partial}
\def\mc{\mathcal}

\usepackage{graphicx}
\usepackage{amsmath}
\usepackage{amssymb}
\usepackage{caption}
\usepackage{subcaption}
\preprint{ \hbox{}\hfill arXiv: 1503.04997}

\title{RG flows from (1,0) 6D SCFTs to N=1 SCFTs in four and three dimensions}
\author{Parinya Karndumri\\
String Theory and Supergravity Group, Department
of Physics, Faculty of Science, Chulalongkorn University, 254 Phayathai Road, Pathumwan, Bangkok 10330, Thailand\\
E-mail: \email{parinya.ka@hotmail.com}}

\abstract{We study $AdS_5\times \Sigma_2$ and $AdS_4\times \Sigma_3$ solutions
of $N=2$, $SO(4)$ gauged supergravity in seven dimensions with
$\Sigma_{2,3}$ being $S^{2,3}$ or $H^{2,3}$. The $SO(4)$ gauged
supergravity is obtained from coupling three vector multiplets to
the pure $N=2$, $SU(2)$ gauged supergravity. With a topological mass
term for the 3-form field, the $SO(4)\sim SU(2)\times SU(2)$ gauged
supergravity admits two supersymmetric $AdS_7$ critical points, with
$SO(4)$ and $SO(3)$ symmetries, provided that the two $SU(2)$ gauge
couplings are different. These vacua correspond to $N=(1,0)$
superconformal field theories (SCFTs) in six dimensions. In the case
of $\Sigma_2$, we find a class of $AdS_5\times S^2$ and $AdS_5\times
H^2$ solutions preserving eight supercharges and $SO(2)\times SO(2)$
symmetry, but only $AdS_5\times H^2$ solutions exist for $SO(2)$
symmetry. These should correspond to some $N=1$ four-dimensional
SCFTs. We also give RG flow solutions from the $N=(1,0)$ SCFTs in
six dimensions to these four-dimensional fixed points including a
two-step flow from the $SO(4)$ $N=(1,0)$ SCFT to the $SO(3)$
$N=(1,0)$ SCFT that eventually flows to the $N=1$ SCFT in four
dimensions. For $AdS_4\times\Sigma_3$, we find a class of
$AdS_4\times S^3$ and $AdS_4\times H^3$ solutions with four
supercharges, corresponding to $N=1$ SCFTs in three dimensions. When
the two $SU(2)$ gauge couplings are equal, only $AdS_4\times H^3$
are possible. The uplifted solutions for equal $SU(2)$ gauge
couplings to eleven dimensions are also given.}


\begin{document}
\section{Introduction}
Six-dimensional superconformal field theories (SCFTs) are interesting in various aspects. In the context of M-theory, these SCFTs arise as a worldvolume theory of M5-branes in the near horizon limit. The correspondence between a six-dimensional $N=(2,0)$ SCFT and M-theory on $AdS_7\times S^4$ is one of the examples given in the AdS/CFT correspondence originally proposed in \cite{maldacena}. This AdS$_7$/CFT$_6$ correspondence has been explored in great details both from the M-theory point of view and the effective $N=4$ $SO(5)$ gauged supergravity in seven dimensions.
\\
\indent In this paper, we are interested in the half-maximal
$N=(1,0)$ SCFTs in six dimensions. It has been shown in
\cite{Seiberg_6D_fixed_points} that $N=(1,0)$ field theory possesses
a non-trivial fixed point, and recently many $N=(1,0)$ SCFTs have
been classified in \cite{6DSCFT_from_Ftheory,Heckman_6DSCFT} and
\cite{Lakshya_6DSCFT}. The holographic study of this $N=(1,0)$
theory has mainly been investigated by orbifolding the $AdS_7\times
S^4$ geometry of eleven-dimensional supergravity, see for example
\cite{Berkooz_6D_dual,AdS7_orbifold1,AdS7_orbifold2}. Recently, many
new $AdS_7$ geometries from massive type IIA string theory have been
found in \cite{All_AdS7}, and the dual SCFTs of these $AdS_7$ vacua
have been studied in \cite{Dual_of_6DN10}.
\\
\indent We are particularly interested in studying $N=(1,0)$ SCFTs
within the framework of seven-dimensional gauged supergravity. These
SCFTs should be dual to $AdS_7$ solutions of $N=2$ gauged
supergravity in seven dimensions \cite{Ferrara_AdS7CFT6}. Pure $N=2$
gauged supergravity with $SU(2)$ gauge group admits both
supersymmetric and non-supersymmetric $AdS_7$ vacua
\cite{Pure_N2_7D1}. The two vacua can be interpreted as a
supersymmetric and a non-supersymmetric CFT, respectively. A domain
wall solution interpolating between these vacua has been studied in
\cite{non_SUSY7Dflow}. This solution describes a non-supersymmetric
deformation of the UV $N=(1,0)$ SCFT to another non-supersymmetric
CFT in the IR.
\\
\indent When coupled to vector multiplets, the $N=2$ gauged
supergravity with many possible gauge groups can be obtained
\cite{Eric_N2_7D,Park_7D,Eric_N2_7Dmassive}. Although the resulting
matter-coupled theory can support only a half-supersymmetric domain
wall vacuum, supersymmetric $AdS_7$ vacua are possible if a
topological mass term for the 3-form field, dual to the 2-form field
in the gravity multiplet, is introduced. These supersymmetric
$AdS_7$ critical points with $SO(4)$ and $SO(3)$ symmetries together
with analytic RG flows interpolating between them have been studied
in \cite{7D_flow} in the case of $SO(4)$ gauge group. And recently,
$AdS_7$ vacua including compactifications to $AdS_5$ of non-compact
gauge groups have been explored in \cite{Non_compact_7DN2}. The
latter type of solutions generally describe twisted
compactifications of $N=(1,0)$ six-dimensional field theories to
four dimensions.
\\
\indent In this paper, we are interested in holographic description
of twisted compactifications of $N=(1,0)$ SCFTs on two-manifolds
$\Sigma_2=(S^2,H^2)$ and three-manifold $\Sigma_3=(S^3,H^3)$. The
corresponding gravity solutions will take the form of $AdS_5\times
\Sigma_2$ and $AdS_4\times \Sigma_3$, respectively. The dual field
theories will be SCFTs in four or three dimensions. Gravity
solutions interpolating between above mentioned $AdS_7$ vacua and
these $AdS_5$ or $AdS_4$ geometries will describe RG flows from
$N=(1,0)$ SCFTs to lower dimensional SCFTs. Previously, this type of
solutions has mainly been studied within the framework of the
maximal $N=4$ gauged supergravity. The solutions provide gravity
duals of twisted compactifications of the $N=(2,0)$ SCFTs. A number
of these $AdS_5$ solutions together with the uplift to eleven-dimensional supergravity by using the
reduction ansatz given in \cite{Peter_S4} and \cite{AdS_BH_embed} have been studied previously in \cite{MN_nogo,Cucu1,Cucu2,Bobev_4DSCFT}. In addition, compactifications of $N=(1,0)$ SCFT has recently been explored from the point of view of massive type IIA theory in \cite{Tomasiello_10SCFT}.
\\
\indent
We will give another new solution to this class from $N=2$ $SO(4)$ gauged supergravity. It has been pointed out in \cite{Cucu1} that the $AdS_5\times S^2$ solution preserving $SO(2)\times SO(2)$ symmetry and $N=2$ supersymmetry in five dimensions, eight supercharges, cannot be obtained from pure minimal $N=2$ gauged supergravity. We will show that this solution is a solution of $N=2$ $SO(4)$ gauged supergravity obtained from coupling pure $N=2$ gauged supergravity to three vector multiplets. We will additionally give new $AdS_5\times H^2$ solutions which are different from those given in \cite{Cucu1} and \cite{Cucu2} in the sense that the two $SU(2)$ gauge couplings are different, and the residual symmetry is only the diagonal subgroup of $SO(2)\times SO(2)$. This case is not a truncation of the $N=4$ $SO(5)$ gauged supergravity, and the embedding of these solutions in higher dimensions are presently unknown. We will also study holographic RG flow solutions interpolating between $AdS_7$ vacua and these $AdS_5$ fixed points. The solutions describe deformations of $N=(1,0)$ SCFTs in six dimensions to the IR $N=1$ SCFT in four dimensions.
\\
\indent On $AdS_4$ solutions from seven-dimensional gauged
supergravity, a class of $AdS_4\times H^3$ and $AdS_4\times S^3$
solutions have been obtained in \cite{Pernici_Sezgin}. A number of
extensive studies of these solutions in terms of wrapped M5-branes
on various supersymmetric cycles in special holonomy manifolds have
been given in \cite{Bobby_AdS4,Wrapped_M5_Kim,Kim_AdS4}. In
particular, the solution studied in \cite{Kim_AdS4} has been
obtained from the maximal gauged supergravity and preserves $N=2$
superconformal symmetry in three dimensions. In this work, we will
look for $AdS_4$ solutions in the $N=2$ $SO(4)$ gauged supergravity
preserving only four supercharges. The corresponding solutions
should then correspond to some $N=1$ SCFTs in three dimensions. We
will show that there exist $AdS_4\times S^3$ and $AdS_4\times H^3$
solutions in this $SO(4)$ gauged supergravity with four supercharges
when the two $SU(2)$ gauge couplings are different. For equal
$SU(2)$ gauge couplings, only $AdS_4\times H^3$ solutions exist and
can be uplifted to eleven dimensions using the reduction ansatz
given in \cite{SO4_7Dfrom11D}.
\\
\indent The paper is organized as follow. In section
\ref{7D_SO4gaugedN2}, relevant information on $N=2$ $SO(4)$ gauged
supergravity in seven dimensions and supersymmetric $AdS_7$ critical
points are reviewed. $AdS_5\times S^2$ and $AdS_5\times H^2$
solutions together with holographic RG flows from $AdS_7$ critical
points to these $AdS_5$ fixed points will be given in section
\ref{AdS5}. We present $AdS_4\times S^3$ and $AdS_4\times H^3$
solutions in section \ref{AdS4} and give the embedding of some
$AdS_5\times \Sigma_2$ and $AdS_4\times \Sigma_3$ solutions in
eleven dimensions in section \ref{uplift11D}. We finally give some
comments and conclusions in section \ref{conclusion}.

\section{Seven-dimensional $N=2$ $SO(4)$ gauged supergravity and $AdS_7$ critical points}\label{7D_SO4gaugedN2}
In this section, we give a description of the $SO(4)$
$N=2$ gauged supergravity in seven dimensions and the associated supersymmetric $AdS_7$ critical points. These critical points preserve $N=2$ supersymmetry in seven dimensions and correspond to six-dimensional $N=(1,0)$ SCFTs. All of the notations used throughout the paper are the same as those in \cite{Eric_N2_7Dmassive} and \cite{7D_flow}.

\subsection{$SO(4)$ gauged supergravity}
The $SO(4)$ $N=2$ gauged supergravity in seven dimensions is constructed by gauging the half-maximal $N=2$ supergravity coupled to three vector multiplets. The supergravity multiplet $(e^m_\mu, \psi^A_\mu,
A^i_\mu,\chi^A,B_{\mu\nu},\sigma)$ consists of the graviton, two gravitini, three vectors, two spin-$\frac{1}{2}$ fields, a two-form field and the dilaton. We will use the convention that curved and flat space-time indices are denoted by $\mu,\nu,\ldots$ and $m,n,\ldots$, respectively. Each vector multiplet $(A_\mu,\lambda^A,\phi^i)$ contains a vector field, two gauginos and three scalars. The bosonic field content of the matter coupled supergravity then consists of the graviton, six vectors and ten scalars parametrized by the $\mathbb{R}^+\times SO(3,3)/SO(3)\times SO(3)\sim \mathbb{R}^+\times SL(4,\mathbb{R})/SO(4)$ coset manifold. In the following, we will consider the supergravity theory in which the two-form field $B_{\mu\nu}$ is dualized to a
three-form field $C_{\mu\nu\rho}$. The latter admits a topological mass term, so the resulting gauged supergravity admits an $AdS_7$ vacuum.
\\
\indent
The $SO(4)$ gauged supergravity is obtained by gauging the $SO(4)\sim SO(3)\times SO(3)$ subgroup of the global symmetry group $SO(3,3)$. One of the $SO(3)$ in the gauge group $SO(3)\times SO(3)$ is the $SO(3)_R\sim USp(2)_R\sim SU(2)_R$ R-symmetry. All spinor fields, including the supersymmetry parameter $\epsilon^A$, are symplectic-Majorana spinors transforming as doublets of the $SU(2)_R$ R-symmetry. From now on, the $SU(2)_R$ douplet indices $A,B=1,2$ will not be shown explicitly. The $SU(2)_R$ triplets are labeled by indices $i,j=1,2,3$ while indices $r,s=1,2,3$ are the triplet indices of the other $SO(3)$ in $SO(3)_R\times SO(3)$.

\indent The $9$ scalar fields in the $SO(3,3)/SO(3)\times SO(3)$ coset are parametrized
by the coset representative $L=(L_I^{\phantom{I}i},L_I^{\phantom{I}r})$ which
transforms under the global $SO(3,3)$ and the local composite $SO(3)\times
SO(3)$ by left and right multiplications, respectively. The inverse of $L$
is denoted by $L^{-1}=(L^I_{\phantom{s}i},L^I_{\phantom{s}r})$ satisfying the relations $L^I_{\phantom{s}i}=\eta^{IJ}L_{Ji}$ and $L^I_{\phantom{s}r}=\eta^{IJ}L_{Jr}$.
\\
\indent The bosonic Lagrangian of the $N=2$ gauged supergravity is given by
\begin{eqnarray}
e^{-1}\mc{L}&=&\frac{1}{2}R-\frac{1}{4}e^\sigma a_{IJ}F^I_{\mu\nu}F^{J\mu\nu}
-\frac{1}{48}e^{-2\sigma}H_{\mu\nu\rho\sigma}H^{\mu\nu\rho\sigma}-\frac{5}{8}\pd_\mu\sigma \pd^\mu\sigma
-\frac{1}{2}P^{ ir}_\mu P^\mu_{ir}\nonumber \\
& &-\frac{1}{144\sqrt{2}}e^{-1}\epsilon^{\mu_1\ldots
\mu_7}H_{\mu_1\ldots\mu_4}
\omega_{\mu_5\ldots\mu_7}+\frac{1}{36}he^{-1}\epsilon^{\mu_1\ldots
\mu_7}H_{\mu_1\ldots\mu_4} C_{\mu_5\ldots\mu_7}-V\nonumber \\
 \label{7Daction}
\end{eqnarray}
where the scalar potential and the Chern-Simons term are given by
\begin{eqnarray}
V&=&\frac{1}{4}e^{-\sigma}\left(C^{ir}C_{ir}-\frac{1}{9}C^2\right)+16h^2e^{4\sigma}
-\frac{4\sqrt{2}}{3}he^{\frac{3\sigma}{2}}C,\\
\omega_{\mu\nu\rho}&=&3\eta_{IJ}F^I_{[\mu\nu}A^J_{\rho]}-f_{IJ}^{\phantom{sa}K}A^I_{\mu}\wedge
A^J_{\nu}\wedge A_{\rho K}
\end{eqnarray}
with the gauge field strength defined by
$F^I_{\mu\nu}=2\pd_{[\mu}A^I_{\nu]}+f_{JK}^{\phantom{sas}I}A^J_{\mu}A^K_{\nu}$. The structure constants $f_{IJ}^{\phantom{sss}K}$ of the gauge group include the gauge coupling associated to each simple factor in a general gauge group $G_0\subset SO(3,3)$.
\\
\indent We are mainly interested in supersymmetric solutions. Therefore, the supersymmetry transformations of fermions are necessary. However, we will not consider bosonic solutions with the three-form field turned on. We will accordingly set $C_{\mu\nu\rho}=0$ throughout. The fermionic supersymmetry transformations, with all fermions and the three-form field vanishing, are given by
\begin{eqnarray}
\delta \psi_\mu &=&2D_\mu
\epsilon-\frac{\sqrt{2}}{30}e^{-\frac{\sigma}{2}}C\gamma_\mu
\epsilon-\frac{i}{20}e^{\frac{\sigma}{2}}F^i_{\rho\sigma}\sigma^i\left(3\gamma_\mu
\gamma^{\rho\sigma}-5\gamma^{\rho\sigma}\gamma_\mu\right)\epsilon
-\frac{4}{5}he^{2\sigma}\gamma_\mu \epsilon,\quad \label{delta_psi}\\
\delta \chi &=&-\frac{1}{2}\gamma^\mu\pd_\mu \sigma
\epsilon-\frac{i}{10}e^{\frac{\sigma}{2}}F^i_{\mu\nu}\sigma^i\gamma^{\mu\nu}\epsilon
+\frac{\sqrt{2}}{30}e^{-\frac{\sigma}{2}}C\epsilon-\frac{16}{5}e^{2\sigma}h\epsilon,\label{delta_chi}\\
\delta \lambda^r &=&-i\gamma^\mu
P^{ir}_\mu\sigma^i\epsilon-\frac{1}{2}e^{\frac{\sigma}{2}}F^r_{\mu\nu}\gamma^{\mu\nu}\epsilon
-\frac{i}{\sqrt{2}}e^{-\frac{\sigma}{2}}C^{ir}\sigma^i\epsilon\, .\label{delta_lambda}
\end{eqnarray}
\indent Various quantities appearing in the Lagrangian and supersymmetry transformations are defined by the following relations
\begin{eqnarray}
D_\mu\epsilon &=&\pd_\mu
\epsilon+\frac{1}{4}\omega_{\mu}^{mn}\gamma_{mn}+\frac{i}{4}\sigma^i\epsilon^{ijk}Q_{\mu
jk},\nonumber \\
P_\mu^{ir}&=&L^{Ir}\left(\delta^K_I\pd_\mu+f_{IJ}^{\phantom{sad}K}A_\mu^J\right)L^i_{\phantom{s}K},
\qquad
Q^{ij}_\mu=L^{Ij}\left(\delta^K_I\pd_\mu+f_{IJ}^{\phantom{sad}K}A_\mu^J\right)L^i_{\phantom{s}K},\nonumber
\\
C_{ir}&=&\frac{1}{\sqrt{2}}f_{IJ}^{\phantom{sad}K}L^I_{\phantom{s}j}L^J_{\phantom{s}k}L_{Kr}\epsilon^{ijk},
\qquad
C=-\frac{1}{\sqrt{2}}f_{IJ}^{\phantom{sad}K}L^I_{\phantom{s}i}L^J_{\phantom{s}j}L_{Kk}\epsilon^{ijk},\nonumber \\
C_{rsi}&=&f_{IJ}^{\phantom{sad}K}L^I_{\phantom{s}r}L^J_{\phantom{s}s}L_{Ki},\qquad
a_{IJ}=L^i_{\phantom{s}I}L_{iJ}+L^r_{\phantom{s}I}L_{rJ},\nonumber \\
F^i_{\mu\nu}&=&L^{\phantom{I}i}_{I}F^I,\qquad
F^r_{\mu\nu}=L^{\phantom{I}r}_{I}F^I
\end{eqnarray}
where $\gamma^m$ are space-time gamma matrices satisfying $\{\gamma^m,\gamma^n\}=2\eta^{mn}$ with $\eta^{mn}=\textrm{diag}(-1,1,1,1,1,1,1)$.

\subsection{Supersymmetric $AdS_7$ critical points}
We will now briefly review supersymmetric $AdS_7$ critical points found in \cite{7D_flow}. There are two critical points preserving the full $N=2$ supersymmetry in seven dimensions. The two critical points however have different symmetries namely one critical point, at which all scalars vanishing, preserves the full $SO(4)$ gauge symmetry while the other is only invariant under the diagonal subgroup $SO(3)_{\textrm{diag}}\subset SO(3)\times SO(3)$.
\\
\indent For $SO(3)\times SO(3)$ gauge group, the gauge structure
constants can be written as \cite{Eric_N2_7Dmassive}
\begin{equation}
f_{IJK}=(g_1\epsilon_{ijk},-g_2\epsilon_{rst}).
\end{equation}
\indent Before discussing the detail of the two critical points, we give an explicit parametrization of the $SO(3,3)/SO(3)\times SO(3)$ coset as follow. With the $36$ basis elements of a general $6\times 6$ matrix
\begin{equation}
(e_{IJ})_{KL}=\delta_{IK}\delta_{JL},\qquad I,J,\ldots =1,\ldots ,6
\end{equation}
the generators of the composite $SO(3)\times SO(3)$ symmetry are given by
\begin{eqnarray}
SO(3)_R&:&\qquad J^{(1)}_{ij}=e_{ji}-e_{ij},\qquad i,j=1,2,3,\nonumber \\
SO(3)&:&\qquad J^{(2)}_{rs}=e_{s+3,r+3}-e_{r+3,s+3},\qquad
r,s=1,2,3\, .
\end{eqnarray}
The non-compact generators corresponding to $9$ scalars take the form of
\begin{equation}
Y^{ir}=e_{i,r+3}+e_{r+3,i}\, .
\end{equation}
Accordingly, the coset representative can be obtained by an exponentiation of the appropriate $Y^{ir}$ generators. $Y^{ir}$ generators and the $9$ scalars transform as $(\mathbf{3},\mathbf{3})$ under the $SO(3)\times SO(3)$ local symmetry.
\\
\indent The supersymmetric $AdS_7$ critical points preserve at least $SO(3)$ symmetry. Therefore, we will consider only the coset representative invariant under $SO(3)$ symmetry. The dilaton $\sigma$ is an $SO(3)\times SO(3)$ singlet. From the $9$ scalars in $SO(3,3)/SO(3)\times SO(3)$, there is one $SO(3)_{\textrm{diag}}$ singlet from the decomposition $\mathbf{3}\times \mathbf{3}\rightarrow \mathbf{1}+\mathbf{3}+\mathbf{5}$. The singlet corresponds to the non-compact generator
\begin{equation}
Y_s=Y^{11}+Y^{22}+Y^{33}\, .\label{SO3_singlet_scalar}
\end{equation}
The coset representative is then given by
\begin{equation}
L=e^{\phi Y_s}\, .\label{SO3_diag_L}
\end{equation}
The scalar potential for the dilaton $\sigma$ and the $SO(3)_{\textrm{diag}}$ singlet scalar $\phi$ can be straightforwardly computed. Its explicit form reads \cite{7D_flow}
\begin{eqnarray}
V&=&\frac{1}{32}e^{-\sigma}\left[(g_1^2+g_2^2)\left(\cosh
(6\phi)-9\cosh(2\phi)\right)+8g_1g_2\sinh^3(2\phi)\phantom{e^{\frac{1}{2}}}\right.\nonumber
\\ & &\left.+8\left[g_2^2-g_1^2+64h^2e^{5\sigma}+32e^{\frac{5\sigma}{2}}h\left(g_1\cosh^2\phi+g_2\sinh^3\phi\right)\right]
\right].
\end{eqnarray}
There are two supersymmetric $AdS_7$ vacua given by
\begin{eqnarray}
SO(4)-\textrm{critical point}&:&\qquad \sigma=\phi=0,\qquad V_0=-240h^2,\label{SO4_AdS7}\\
SO(3)-\textrm{critical point}&:&\qquad \sigma=-\frac{1}{5}\ln
\left[\frac{g_2^2-256h^2}{g_2^2}\right],\nonumber \\
\phi &=&\frac{1}{2}\ln\left[\frac{g_2+16h}{g_2-16h}\right], \qquad
V_0=-\frac{240g_2^{\frac{8}{5}}h^2}{(g_2^2-256h^2)^{\frac{4}{5}}}\label{SO3_AdS7}
\end{eqnarray}
where we have chosen $g_1=-16h$ in order to make the $SO(4)$ critical point occurs at $\sigma=0$. This is achieved by shifting $\sigma$. The value of the cosmological constant has been denoted by $V_0$.
\\
\indent The two critical points correspond to $N=(1,0)$ SCFTs in six dimensions with $SO(4)$ and $SO(3)$ symmetries, respectively. An RG flow solution interpolating between these two critical points has already been studied in \cite{7D_flow}. In the next sections, we will study supersymmetric RG flows from these SCFTs to other SCFTs in four and three dimensions providing holographic descriptions of twisted compactifications of these $N=(1,0)$ SCFTs.

\section{Flows to $N=1$ SCFTs in four dimensions}\label{AdS5}
In this section, we look for solutions of the form $AdS_5\times S^2$ or $AdS_5\times H^2$ in which $S^2$ and $H^2$ are a two-sphere and a two-dimensional hyperbolic space, respectively.
\\
\indent In the case of $S^2$, we take the seven-dimensional metric to be
\begin{equation}
ds^2_7=e^{2F(r)}dx_{1,3}^2+dr^2+e^{2G(r)}(d\theta^2+\sin^2d\phi^2)\label{AdS5S2_metric}
\end{equation}
with $dx^2_{1,3}$ being the flat metric on the four-dimensional spacetime. By using the vielbein
\begin{eqnarray}
e^{\hat{\mu}}&=&e^Fdx^\mu,\qquad e^{\hat{r}}=dr,\nonumber \\
e^{\hat{\theta}}&=&e^{G}d\theta ,\qquad e^{\hat{\phi}}=e^G\sin \theta d\phi,
\end{eqnarray}
we can compute the following spin connections
\begin{eqnarray}
\omega^{\hat{\phi}}_{\phantom{s}\hat{\theta}} &=&e^{-G}\cot\theta e^{\hat{\phi}},\qquad
\omega^{\hat{\phi}}_{\phantom{s}\hat{r}}=G'e^{\hat{\phi}},\nonumber \\
\omega^{\hat{\theta}}_{\phantom{s}\hat{r}}&=&G'e^{\hat{\theta}},\qquad \omega^{\hat{\mu}}_{\phantom{s}\hat{r}}=F'e^{\hat{\mu}}\, .
\end{eqnarray}
where $'$ denotes the $r$-derivative. Hatted indices are tangent space indices.
\\
\indent In the case of $H^2$, we take the matric to be
\begin{equation}
ds^2_7=e^{2F(r)}dx^2_{1,3}+dr^2+\frac{e^{2G(r)}}{y^2}(dx^2+dy^2).\label{metric_AdS5_H2}
\end{equation}
With the vielbein
\begin{eqnarray}
e^{\hat{\mu}}&=&e^Fdx^\mu ,\qquad e^{\hat{r}}=dr,\nonumber \\
e^{\hat{x}}&=&\frac{e^G}{y}dx,\qquad e^{\hat{y}}=\frac{e^G}{y}dy,
\end{eqnarray}
the spin connections are found to be
\begin{eqnarray}
\omega^{\hat{x}}_{\phantom{s}\hat{r}}&=&G'e^{\hat{x}},\qquad
\omega^{\hat{y}}_{\phantom{s}\hat{r}}=G'e^{\hat{y}},\nonumber \\
\omega^{\hat{\mu}}_{\phantom{s}\hat{r}}&=&F'e^{\hat{\mu}},\qquad
\omega^{\hat{x}}_{\phantom{s}\hat{y}}=-e^{-G(r)}e^{\hat{x}}\, .\label{spin_connection_H2}
\end{eqnarray}

\subsection{$AdS_5$ solutions with $SO(2)\times SO(2)$ symmetry}
We now construct the BPS equations from the supersymmetry transformations of fermions. We first consider the $S^2$ case. In order to preserve supersymmetry, we make a twist by turning on the $SO(2)\times SO(2)\subset SO(4)$ gauge fields, among the six gauge fields $A^I$,
\begin{equation}
A^3=a\cos \theta  d\phi \qquad\textrm{and}\qquad A^6=b\cos\theta d\phi\label{A3_A6}
\end{equation}
such that the spin connections on $S^2$ is cancelled by these gauge connections. The Killing spinor corresponding to the unbroken supersymmetry is then a constant spinor on $S^2$.
\\
\indent We begin with the solutions preserving the full $SO(2)\times SO(2)$ residual gauge symmetry generated by $J^{(1)}_{12}$ and $J^{(2)}_{12}$. Scalars which are singlet under $SO(2)\times SO(2)$ are the dilaton and the scalar corresponding to the $SO(3,3)$ non-compact generators $Y^{33}$. We will denote this scalar by $\Phi$. By considering the variation of the gravitino along $S^2$ directions, we find that the cancellation between the spin and gauge connections imposes the twist condition
\begin{equation}
ag_1=1\, .\label{S2_twist}
\end{equation}
\indent Using the projection conditions
\begin{equation}
\gamma_r\epsilon =\epsilon,\qquad \textrm{and}\qquad i\sigma^3 \gamma^{\hat{\theta}\hat{\phi}}\epsilon=\epsilon,
\end{equation}
we find the following BPS equations
\begin{eqnarray}
\Phi'&=&\frac{1}{2}e^{-\frac{\sigma}{2}-\Phi-2G}\left[e^{2G}g_1(e^{2\Phi}-1)-ae^\sigma(e^{2\Phi}-1)
-be^{\sigma}(e^{2\Phi}+1)\right],
\label{S2_eq1}\\
\sigma'&=&\frac{1}{5}e^{-\frac{\sigma}{2}-\Phi-2G}\left[e^\sigma\left[a-b+(a+b)e^{2\Phi}\right]-e^{2G}
\left(g_1+g_1e^{2\Phi}+32he^{\frac{5\sigma}{2}+\Phi}\right)\right],\qquad\label{S2_eq2}\\
G'&=&-\frac{1}{10}e^{-\frac{\sigma}{2}-\Phi-2G}\left[4e^{\sigma}\left[a-b+(a+b)e^{2\Phi}\right]+e^{2G}\left(g_1+g_1e^{2\Phi}
-8he^{\frac{5\sigma}{2}+\Phi}\right)\right],\qquad \label{S2_eq3}\\
F'&=&\frac{1}{10}e^{-\frac{\sigma}{2}-\Phi-2G}\left[e^{\sigma}\left[a-b+(a+b)e^{2\Phi}\right]-e^{2G}\left(g_1+g_1e^{2\Phi}
-8he^{\frac{5\sigma}{2}+\Phi}\right)\right].\label{S2_eq4}
\end{eqnarray}
\\
\indent In the $H^2$ case, we choose the gauge fields to be
\begin{equation}
A^3=\frac{a}{y}dx\qquad \textrm{and}\qquad A^6=\frac{b}{y}dx
\end{equation}
which can be verified that the spin connection
$\omega^{\hat{x}\hat{y}}$ in \eqref{spin_connection_H2} is cancelled
by virtue of the twist condition \eqref{S2_twist} and the projection
conditions
\begin{equation}
\gamma_r \epsilon=\epsilon\qquad \textrm{and} \qquad
i\sigma^3\gamma^{\hat{x}\hat{y}}\epsilon=\epsilon\,
.\label{H2_projector}
\end{equation}
By an analogous computation, we find a similar set of BPS equations
as in \eqref{S2_eq1}, \eqref{S2_eq2}, \eqref{S2_eq3} and
\eqref{S2_eq4} with $(a,b)$ replaced by $(-a,-b)$.
\\
\indent At large $r$, solutions to the above BPS equations should approach the $SO(4)$
$AdS_7$ critical point with $\Phi\sim \sigma\sim 0$ and $F\sim G\sim
r$. This is the UV $(1,0)$ SCFT. As $r\rightarrow -\infty$, we look
for the solution of the form $AdS_5\times S^2$ or $AdS_5\times H^2$ such that
$\phi'=\sigma'=G'=0$ and $F'=\textrm{constant}$. We find that there
is an $AdS_5$ solution given by
\begin{eqnarray}
\Phi &=&\frac{1}{2}\ln \left[\frac{b\pm
\sqrt{4a^2-3b^2}}{2(a+b)}\right],\nonumber \\
\sigma &=&\frac{1}{5}\ln \left[\frac{g_1^2b^2(b\pm \sqrt{4a^2-3b^2})}{32(a+b)h^2(3b-2a\pm \sqrt{4a^2-3b^2})}
\right],\nonumber \\
G&=&\frac{1}{10}\ln
\left[\frac{b^2(a+b)^4(b\pm\sqrt{4a^2-3b^2})(2a-3b\mp
\sqrt{4a^2-3b^2})^3}{32g_1^3h^2(2a+b\mp\sqrt{4a^2-3b^2})^5}\right],\nonumber
\\
L_{AdS_5}&=&\left[\frac{(a+b)^2(2a-3b\pm\sqrt{4a^2-3b^2})^4}
{b^4g_1^4h(b\mp\sqrt{4a^2-3b^2})^2}\right]^{\frac{1}{5}}\,
.\label{SO2SO2_AdS5}
\end{eqnarray}
This solution is given for $\Sigma_2=S^2$. The solution in the $H^2$ case is given similarly by flipping the signs of $a$ and $b$.
\\
\indent It should be noted that, in this fixed point solution with
$SO(2)\times SO(2)$ symmetry, the coupling $g_2$ does not appear.
The solution can then be taken as a solution of the gauged supergravity with
$g_2=g_1$. Therefore, the solution can be uplifted to eleven
dimensions by using the reduction ansatz in \cite{SO4_7Dfrom11D}.
This will be done in section \ref{uplift11D}. The uplifted solution
is however not new since similar solutions have been found previously
in \cite{Cucu1,Cucu2}, and supergravity solutions interpolating between
$AdS_7$ and $AdS_5\times S^2$ or $AdS_5\times H^2$ have also been investigated.
The solutions have an interpretation in terms of RG flows from the
UV SCFT in six dimensions to four-dimensional SCFTs with
$SO(2)\times SO(2)$ symmetry.
\\
\indent Note also that, in this case, it is not possible to find an
RG flow from the $SO(3)$ $AdS_7$ point to any of these
four-dimensional SCFTs since this $AdS_7$ critical point is not
accessible from the BPS equations given above.

\subsection{$AdS_5$ solutions with $SO(2)$
symmetry}\label{SO2_DAdS5}
We now consider $AdS_5$ solutions with $SO(2)$ symmetry. We will
study two possibilities namely the $SO(2)_{\textrm{diag}}\subset
SO(2)\times SO(2)\subset SO(3)\times SO(3)$ and $SO(2)_R\subset
SO(3)_R$.

\subsubsection{Flows with $SO(2)_{\textrm{diag}}$ symmetry}\label{SO2_DAdS51}
\begin{figure}
         \centering
         \begin{subfigure}[b]{0.3\textwidth}
                 \includegraphics[width=\textwidth]{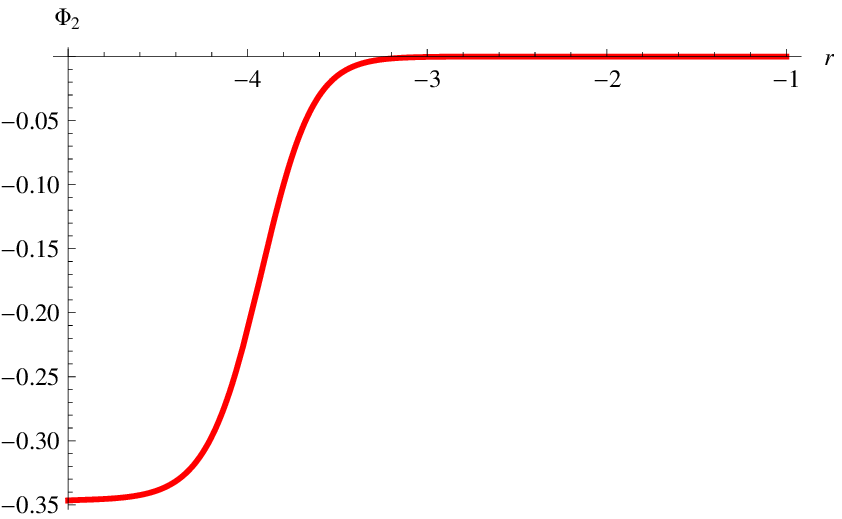}
                 \caption{$\Phi_2$ solution}
         \end{subfigure}%
         \begin{subfigure}[b]{0.3\textwidth}
                 \includegraphics[width=\textwidth]{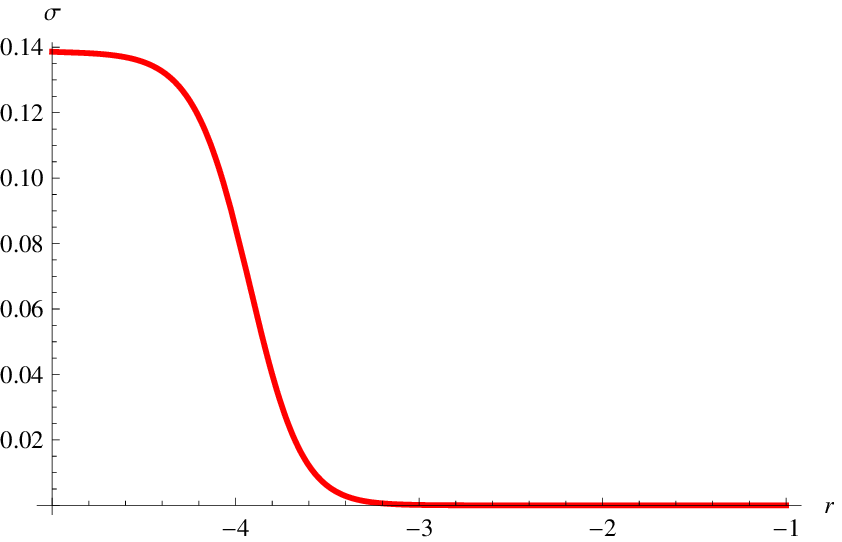}
                 \caption{$\sigma$ solution}
         \end{subfigure}%
         \begin{subfigure}[b]{0.3\textwidth}
                 \includegraphics[width=\textwidth]{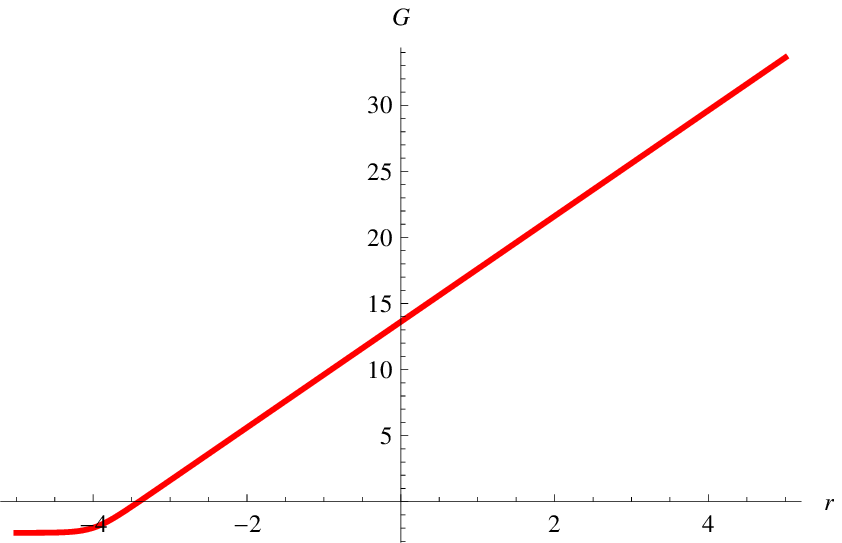}
                 \caption{$G$ solution}
         \end{subfigure}
         \caption{RG flows from $SO(4)$ $N=(1,0)$ SCFT in six dimensions to four-dimensional
         $N=1$ SCFT with $SO(2)_{\textrm{diag}}$ symmetry for $g_1=g_2$.}\label{fig1}
 \end{figure}

\begin{figure}
         \centering
         \begin{subfigure}[b]{0.3\textwidth}
                 \includegraphics[width=\textwidth]{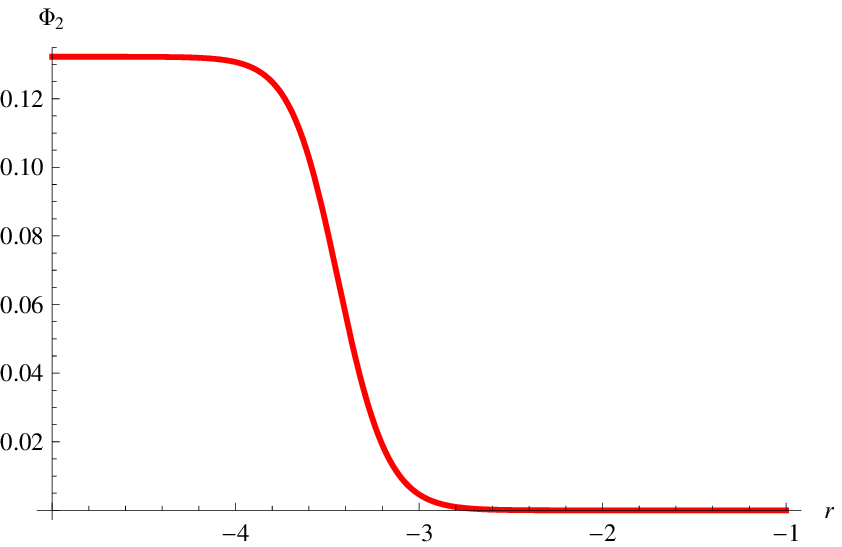}
                 \caption{$\Phi_2$ solution}
         \end{subfigure}%
         \begin{subfigure}[b]{0.3\textwidth}
                 \includegraphics[width=\textwidth]{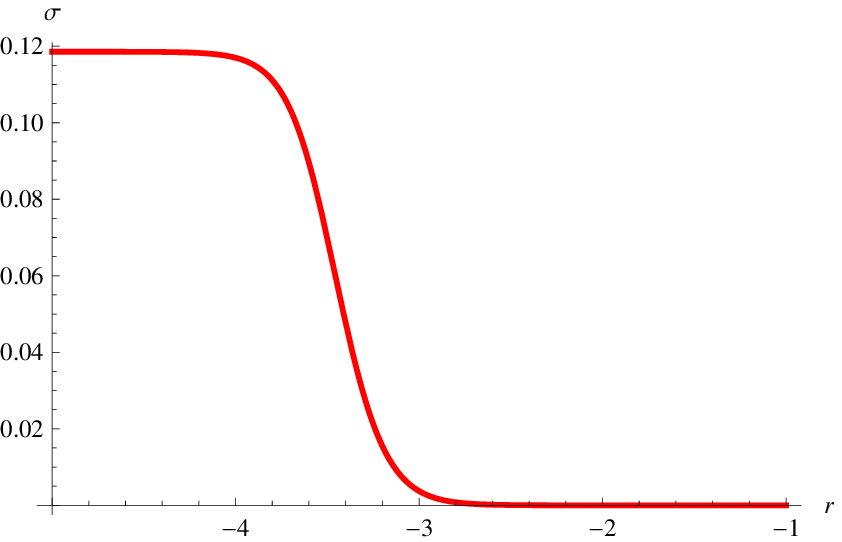}
                 \caption{$\sigma$ solution}
         \end{subfigure}%
         \begin{subfigure}[b]{0.3\textwidth}
                 \includegraphics[width=\textwidth]{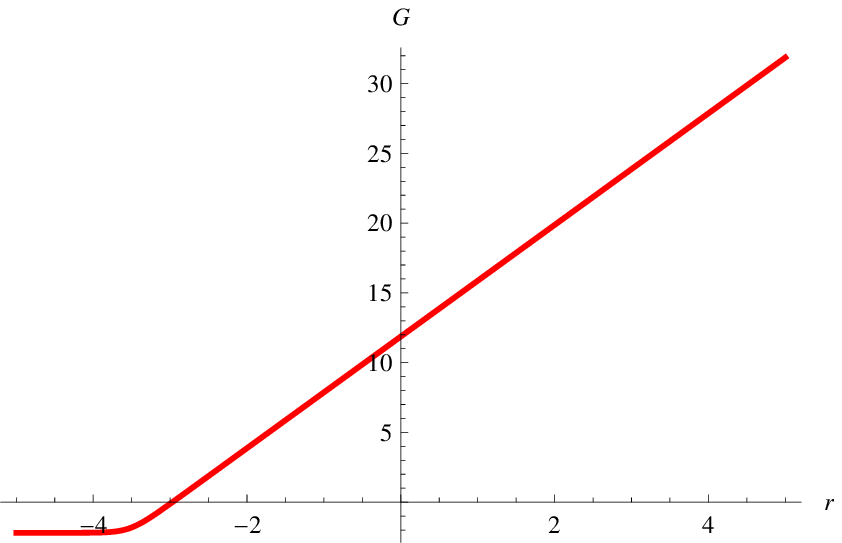}
                 \caption{$G$ solution}
         \end{subfigure}
         \caption{RG flows from $SO(4)$ $N=(1,0)$ SCFT in six dimensions to four-dimensional
         $N=1$ SCFT with $SO(2)_{\textrm{diag}}$ symmetry for $g_1\neq g_2$.}\label{fig2}
 \end{figure}

\begin{figure}
         \centering
         \begin{subfigure}[b]{0.4\textwidth}
                 \includegraphics[width=\textwidth]{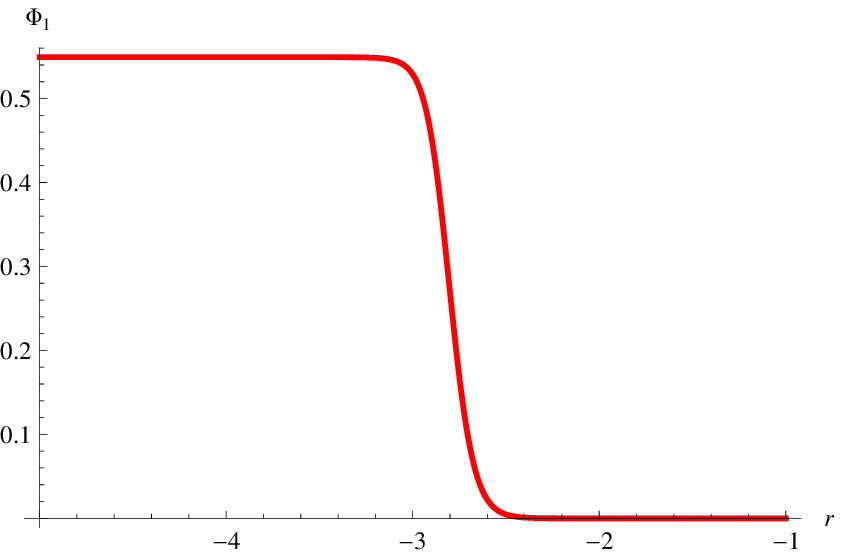}
                 \caption{$\Phi_1$ solution}
         \end{subfigure}%
         \begin{subfigure}[b]{0.4\textwidth}
                 \includegraphics[width=\textwidth]{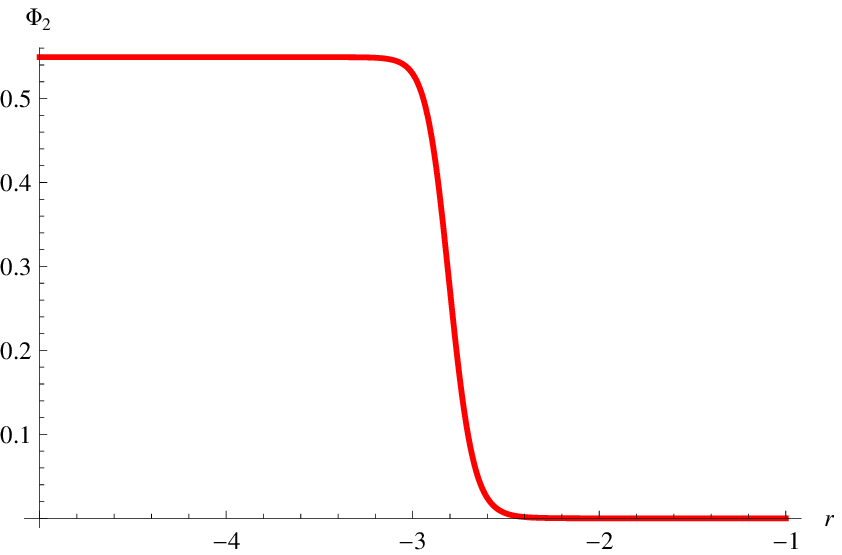}
                 \caption{$\Phi_2$ solution}
         \end{subfigure}\\%
         ~ 
         \begin{subfigure}[b]{0.4\textwidth}
                 \includegraphics[width=\textwidth]{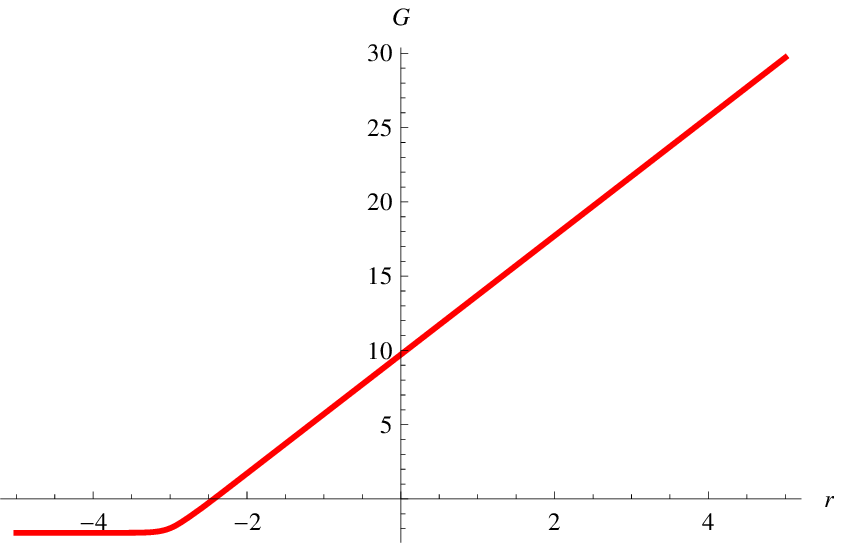}
                 \caption{$G$ solution}
         \end{subfigure}
         \begin{subfigure}[b]{0.4\textwidth}
                 \includegraphics[width=\textwidth]{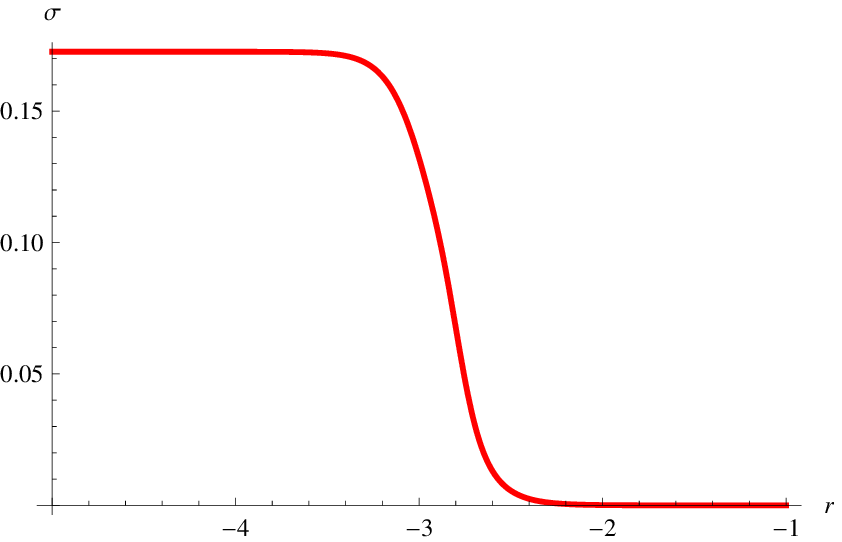}
                 \caption{$\sigma$ solution}
         \end{subfigure}%
         \caption{An RG flow from $SO(4)$ $N=(1,0)$ SCFT in six dimensions to four-dimensional
         $N=1$ SCFT with $SO(2)_{\textrm{diag}}$ symmetry.}\label{fig3}
 \end{figure}

\begin{figure}
         \centering
         \begin{subfigure}[b]{0.4\textwidth}
                 \includegraphics[width=\textwidth]{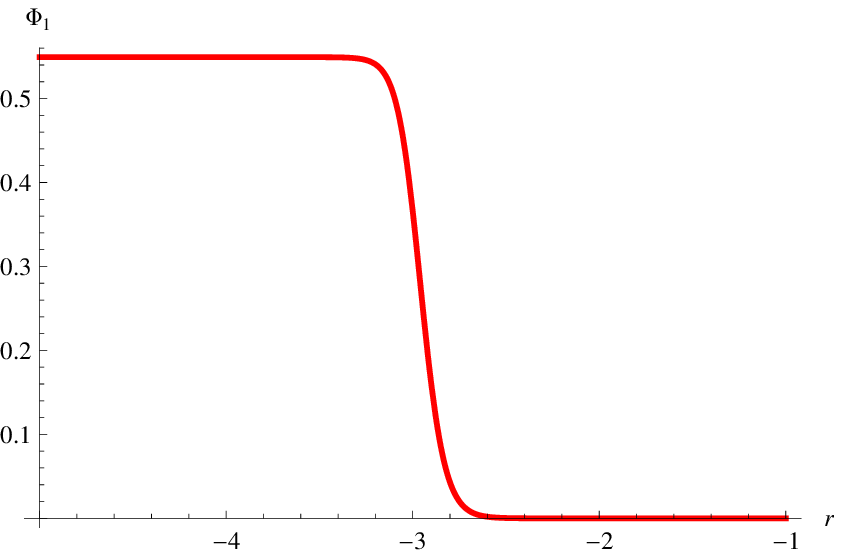}
                 \caption{$\Phi_1$ solution}
         \end{subfigure}%
         \begin{subfigure}[b]{0.4\textwidth}
                 \includegraphics[width=\textwidth]{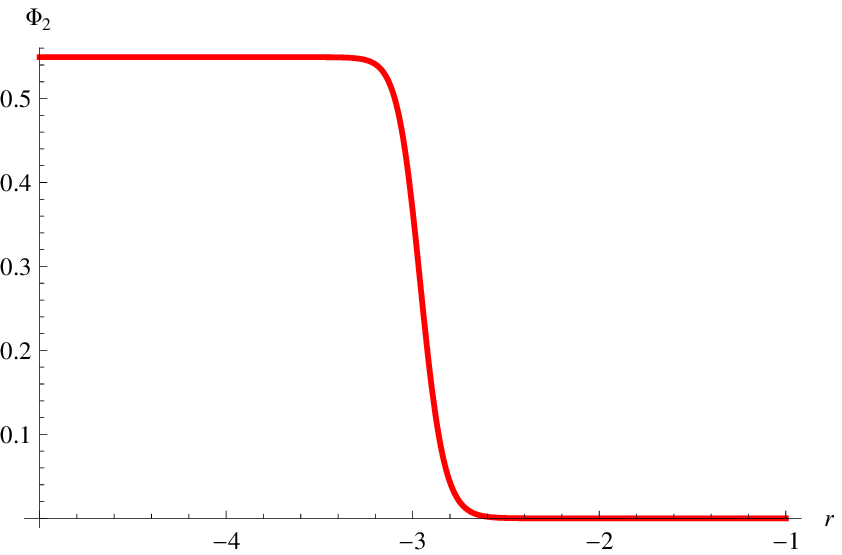}
                 \caption{$\Phi_2$ solution}
         \end{subfigure}\\%
         ~ 
         \begin{subfigure}[b]{0.4\textwidth}
                 \includegraphics[width=\textwidth]{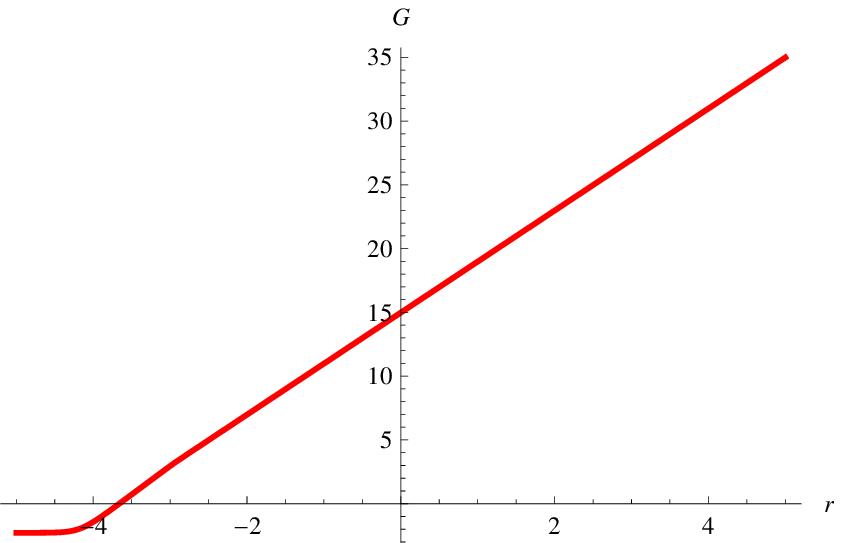}
                 \caption{$G$ solution}
         \end{subfigure}
         \begin{subfigure}[b]{0.4\textwidth}
                 \includegraphics[width=\textwidth]{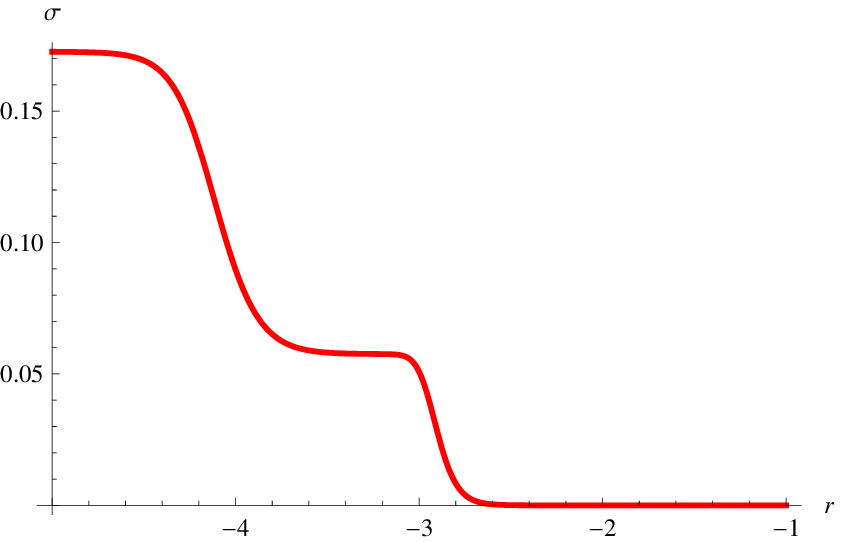}
                 \caption{$\sigma$ solution}
         \end{subfigure}%
         \caption{An RG flow from $SO(4)$ $N=(1,0)$ SCFT to $SO(3)$ $N=(1,0)$ SCFT
         in six dimensions and then to $N=1$ four-dimensional SCFT with $SO(2)_{\textrm{diag}}$ symmetry.}\label{fig4}
 \end{figure}

We begin with the $SO(2)_{\textrm{diag}}$ symmetry generated by $J^{(1)}_{12}+J^{(2)}_{12}$. Among the $9$ scalars in $SO(3,3)/SO(3)\times SO(3)$, there are three singlets under $SO(2)_{\textrm{diag}}$ corresponding to the following decomposition of $SO(3)\times SO(3)$ representations under $SO(2)_{\textrm{diag}}$
\begin{equation}
\mathbf{3}\times \mathbf{3}=(\mathbf{2}+\mathbf{1})\times (\mathbf{2}+\mathbf{1})=\mathbf{1}+\mathbf{1}+\mathbf{2}+\mathbf{2}+\mathbf{2}+\mathbf{1}\, .
\end{equation}
The three singlets correspond to the non-compact generators
\begin{equation}
Y^{11}+Y^{22},\qquad Y^{33},\qquad Y^{12}-Y^{21}\, .
\end{equation}
The coset representative describing these singlets can be written as
\begin{equation}
L=e^{\Phi_1 (Y^{11}+Y^{22})}e^{\Phi_2 Y^{33}}e^{\Phi_3 (Y^{12}-Y^{21})}\, .
\end{equation}
\indent Since we have not found any $AdS_5\times S^2$ solution, we
will give only the result for the $H^2$ case. The $SO(2)_{\textrm{diag}}$ gauge field can be obtained from the $SO(2)\times SO(2)$ gauge fields in \eqref{A3_A6} with the condition that
\begin{equation}
bg_2=ag_1\, .
\end{equation}
As in the previous case, the twist imposes the condition $g_1a=1$ which in the present case also implies $g_2b=1$.
\\
\indent Using the projection conditions \eqref{H2_projector}, we
find the following BPS equations
\begin{eqnarray}
\Phi_1'&=&\frac{1}{8}e^{-\frac{\sigma}{2}-2\Phi_1-\Phi_2}(e^{4\Phi_1}-1)
\left[g_1-g_2+(g_1+g_2)e^{2\Phi_2}\right],\\
\Phi_2'&=&\frac{1}{16g_2}e^{-\frac{\sigma}{2}}
\left[8g_1a\left[g_1-g_2+(g_1+g_2)e^{\Phi_2}\right]\right.\nonumber \\
&
&+g_2\left[e^{-2\Phi_1-\Phi_2-2\Phi_3}(1+e^{4\Phi_1})(1+e^{4\Phi_3})
\left[g_2-g_1+(g_1+g_2)e^{2\Phi_2}\right] \right.\nonumber \\
& &\left.\left.+4(g_1-g_2)e^{\Phi_2}-(g_1+g_2)e^{-\Phi_2}\right]\right],\\
\Phi_3'&=&\frac{1}{8}e^{-\frac{\sigma}{2}-\Phi_2-2\Phi_3}(e^{4\Phi_3}-1)
\left[g_1-g_2+(g_1+g_2)e^{2\Phi_2}\right],\\
\sigma'&=&\frac{1}{40g_2}e^{-\frac{\sigma}{2}-2\Phi_1-\Phi_2-2\Phi_3}\left[8ae^{\sigma+2\Phi_1+2\Phi_3-2G}
\left[g_1-g_2-(g_1+g_2)e^{2\Phi_2}\right]\right.\nonumber \\
&
&-g_2\left[g_1(1+e^{2\Phi_2})(1+e^{4\Phi_1}+e^{4\Phi_3}+4e^{2\Phi_1+2\Phi_3}+e^{4\Phi_1+4\Phi_3})
\right.\nonumber \\
& &
+g_2(e^{2\Phi_2}-1)(1+e^{4\Phi_1}+e^{4\Phi_3}-4e^{2\Phi_1+2\Phi_3}+e^{4\Phi_1+4\Phi_3})\nonumber \\
& &
\left.\left.+256he^{\frac{5\sigma}{2}+2\Phi_1+\Phi_2+2\Phi_3}
\right] \right],
\\
G'&=&\frac{1}{20}e^{-\frac{\sigma}{2}}\left[16he^{\frac{5\sigma}{2}}-g_1(e^{\Phi_2}+e^{-\Phi_2})
+g_2(e^{\Phi_2}-e^{-\Phi_2})\right.\nonumber \\
&
&-\frac{1}{4}e^{-2\Phi_1-\Phi_2-2\Phi_3}(1+e^{4\Phi_1})(1+e^{4\Phi_3})[g_1-g_2+(g_1+g_2)e^{2\Phi_2}]
\nonumber\\
&
&\left.+\frac{8a}{g_2}e^{\sigma-\Phi_2-2G}[g_2-g_1+(g_1-g_2)e^{2\Phi_2}]
\right],\\
F'&=&\frac{1}{20}e^{-\frac{\sigma}{2}}\left[16he^{\frac{5\sigma}{2}}-g_1(e^{\Phi_2}+e^{-\Phi_2})
+g_2(e^{\Phi_2}-e^{-\Phi_2})\right.\nonumber \\
&
&-\frac{1}{4}e^{-2\Phi_1-\Phi_2-2\Phi_3}(1+e^{4\Phi_1})(1+e^{4\Phi_3})[g_1-g_2+(g_1+g_2)e^{2\Phi_2}]
\nonumber\\
&
&\left.-\frac{2a}{g_2}e^{\sigma-\Phi_2-2G}[g_2-g_1+(g_1-g_2)e^{2\Phi_2}]
\right].
\end{eqnarray}
In this case, there are a number of possible $AdS_5$ fixed point solutions, and it is possible to have a solution
interpolating between the $SO(3)$ $AdS_7$ critical points and the
$AdS_5$ in the IR. We will investigate each of them in the following discussion.
\\
\indent We first look at the $AdS_5\times H^2$ critical point with
$g_2=g_1$ since this can be uplifted to eleven dimensions. When
$g_2=g_1$, the fixed point solution exists only for
$\Phi_1=\Phi_3=0$, and the corresponding solution is given by
\begin{eqnarray}
\Phi_2&=&-\frac{1}{2}\ln 2, \sigma =\frac{1}{5}\ln 2,\nonumber \\
G&=&\frac{3}{5}\ln 2-\frac{1}{2}\ln \left[\frac{g_1}{a}\right],
\qquad
L_{AdS_5}=\frac{1}{2^{\frac{12}{5}}h}\label{SO2diag_AdS5_g2_equal_g1}
\end{eqnarray}
The $AdS_5$ solution preserves eight supercharges corresponding to
$N=1$ superconformal field theory in four dimensions with $SO(2)$
symmetry. A flow solution interpolating between this $AdS_5\times H^2$
fixed point and the $SO(4)$ $AdS_7$ given in \eqref{SO4_AdS7} for
$h=1$ is shown in Figure \ref{fig1}.
\\
\indent It should be noted here that this fixed point can be
obtained from the $SO(2)\times SO(2)$ fixed points given in the
previous section by setting the
parameter $b=a$. It can be readily verified that, for $b=a$, solution in
\eqref{SO2SO2_AdS5} is valid only for the upper sign and
$\Sigma_2=H^2$. The resulting solution is precisely that given in
\eqref{SO2diag_AdS5_g2_equal_g1}.
\\
\indent We now move to solutions with $g_2\neq g_1$. The solution
given in \eqref{SO2diag_AdS5_g2_equal_g1} is a special case of a
more general solution, with $\Phi_1=\Phi_3=0$ and $g_2\neq g_1$, which is given by
\begin{eqnarray}
\Phi_2&=&\frac{1}{2}\ln \left[\frac{g_1\pm
\sqrt{4g_2^2-3g_1^2}}{2(g_1+g_2)}\right],\qquad
\Phi_1=\Phi_3=0,\nonumber \\
\sigma &=&\frac{1}{5}\left[\frac{1024h^2(\sqrt{g_2^2-192h^2}\mp
8h)}{(g_2-16h)(g_2+24h\mp \sqrt{g_2^2-192h^2})}\right],\nonumber \\
G&=&\frac{1}{10}\ln
\left[\frac{a^5(g_2-16h)^4(\sqrt{g_2^2-192h^2}\mp 8h)(g_2+24h\mp
\sqrt{g_2^2-192h^2})^3}{1024g_2^5h^3(g_2-8h\mp
\sqrt{g_2^2-192h^2})^5}\right],\nonumber \\
L_{AdS_5}&=&\frac{1}{2}\left[\frac{(g_2-16h)^2(g_2+24h\mp
\sqrt{g_2^2-192h^2})^4}{2h^9(8h\mp
\sqrt{g_2^2-192h^2})}\right]^{\frac{1}{5}}
\end{eqnarray}
where we have used the relation $g_1=-16h$ in the solutions for
$\sigma$ and $G$ to simplify the expressions. An example of the
corresponding flow solutions from the UV $N=(1,0)$ $SO(4)$ SCFT to
this critical point, with $g_2=-2g_1$ and $h=1$, is given in Figure
\ref{fig2}.
\\
\indent In all of the above solutions, it is not possible to have a
flow from the $SO(3)$ $AdS_7$ critical point \eqref{SO3_AdS7}. To
find this type of flows, we look for $AdS_5$ fixed points with
$\Phi_3=0$ but $\Phi_1\neq 0$ and $\Phi_2\neq 0$. In this case, the
$AdS_5\times H^2$ solution is given by
\begin{eqnarray}
\Phi_2&=&\frac{1}{2}\ln \left[\frac{g_2-g_1}{g_2+g_1}\right],\qquad
\Phi_1=\pm \Phi_2,\nonumber \\
\sigma &=&\frac{1}{5}\ln
\left[\frac{g_1^2g_2^2}{144h^2(g_2^2-g_1^2)}\right],\qquad
G=\frac{1}{2}\ln
\left[\frac{2^{\frac{4}{5}}(3^{\frac{3}{5}})a(g_2^2-256h^2)^{\frac{4}{5}}}{g_2^{\frac{8}{5}}g_1}\right],
\nonumber\\
L_{AdS_5}&=&\frac{3^{\frac{4}{5}}(g_2^2-256h^2)^{\frac{2}{5}}}{2^{\frac{18}{5}}g_2^{\frac{4}{5}}h}\,
.\label{SO2diag_AdS5}
\end{eqnarray}
\indent Note that at the values of $\Phi_1$ and $\Phi_2$ are the
same as the $SO(3)$ $AdS_7$ point. In equation \eqref{SO3_AdS7}, we
have
\begin{equation}
\Phi_1=\Phi_2=\frac{1}{2}\ln
\left[\frac{g_2-g_1}{g_2+g_1}\right]\equiv \Phi_0\, .
\end{equation}
Actually, there are two
equivalent values of $\Phi_1$ namely either $\Phi_1=\Phi_0$ or $\Phi_1=-\Phi_0$. The two choices are
equivalent in the sense that they give rise to the same value of the
cosmological constant and the same scalar masses. The difference
between the two is the generators of $SO(3)$ under which the $SO(3)$
singlet scalar $\phi$ in \eqref{SO3_AdS7} is invariant. For
$\Phi_1=\Phi_0$, we
have $\Phi_1=\Phi_2$ which is invariant under the $SO(3)$ generated
by $J^{(1)}_{ij}+J^{(2)}_{ij}$. The alternative value of
$\Phi_1=-\Phi_0$ gives
$\Phi_1=-\Phi_2$ which is invariant under $SO(3)$ generators
$J^{(1)}_{12}+J^{(2)}_{12}$, $J^{(1)}_{13}-J^{(2)}_{13}$ and
$J^{(1)}_{23}-J^{(2)}_{23}$. This difference does not affect the
result discussed here since, in both cases, the residual
$SO(2)_{\textrm{diag}}$ is still generated by
$J^{(1)}_{12}+J^{(2)}_{12}$.
\\
\indent The flow from $SO(3)$ $N=(1,0)$ SCFT would be driven only by the
dilaton $\sigma$ which has different values at the $SO(3)$ $AdS_7$ and the
$AdS_5$ fixed points. This is expected since at $SO(3)$ $AdS_7$
critical point only $\sigma$ corresponds to relevant operators, see
the scalar masses in \cite{7D_flow}.
\\
\indent We now consider RG flows from $N=(1,0)$ SCFTs in six
dimensions to four-dimensional SCFTs identified with the
critical point \eqref{SO2diag_AdS5}. In order to give some explicit
examples, we choose particular values of the two couplings $g_1$
and $g_2$. In the following solutions, we will set $g_2=-2g_1$ and
$h=1$. With these, the IR $AdS_5\times H^2$ is
given by
\begin{eqnarray}
\Phi_1&=&\Phi_2=\frac{1}{2}\ln 3\approx 0.5493,\qquad \sigma=\frac{1}{5}\ln \left[\frac{64}{27}\right]
\approx 0.1726,\nonumber \\
G&=&\frac{1}{10}\ln \left[\frac{3^{7}}{2^{44}}\right]\approx
-2.2808\, .
\end{eqnarray}
The $SO(4)$ UV point \eqref{SO4_AdS7} is given by
\begin{equation}
\Phi_1=\Phi_2=\sigma=0
\end{equation}
while the $SO(3)$ $AdS_7$ point \eqref{SO3_AdS7} occurs at
\begin{equation}
\sigma=\frac{1}{5}\ln\frac{4}{3}\approx 0.0575,\qquad \Phi_2=\Phi_1=\frac{1}{2}\ln 3\approx 0.5493\, .
\end{equation}
We have chosen $\Phi_1=\Phi_2$ at the IR fixed points for definiteness.
\\
\indent There exist an RG flow from the $SO(4)$ $N=(1,0)$ SCFT in
the UV to the $N=1$ four-dimensional SCFT in the IR as shown in
Figure \ref{fig3}. With a particular boundary condition, we can find
an RG flow from the $SO(4)$ $AdS_7$ to the $SO(3)$ $AdS_7$ critical
points and then to the $AdS_5$ critical point as shown in Figure
\ref{fig4}. This solution is similar to the flow from $SO(6)$
$AdS_5$ to Khavaev-Pilch-Warner (KPW) $AdS_5$ critical point and continue to
a two-dimensional $N=(2,0)$ SCFT in \cite{Bobev_4D2D_flow}.

\subsubsection{Flows with $SO(2)_R$ symmetry}\label{SO2R_flow}
\begin{figure}
         \centering
         \begin{subfigure}[b]{0.4\textwidth}
                 \includegraphics[width=\textwidth]{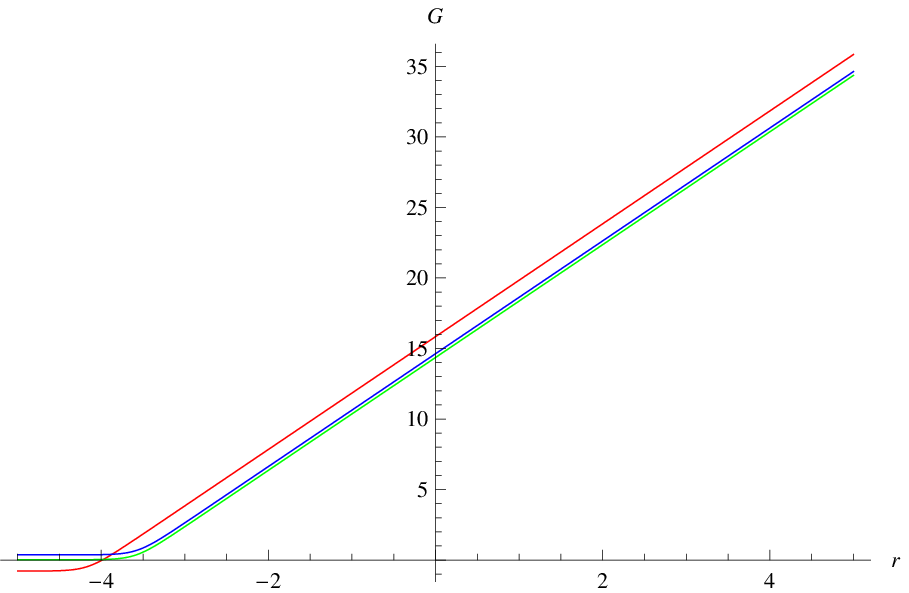}
                 \caption{$\Phi$ solution}
         \end{subfigure}%
         \begin{subfigure}[b]{0.4\textwidth}
                 \includegraphics[width=\textwidth]{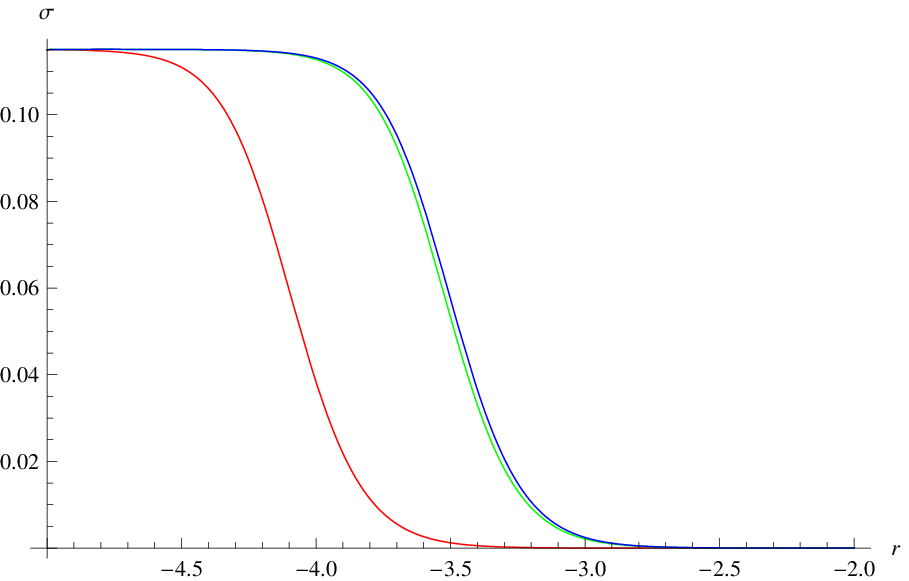}
                 \caption{$\sigma$ solution}
         \end{subfigure}%
         \caption{RG flows from $SO(4)$ $N=(1,0)$ SCFT in six dimensions to four-dimensional
         $N=1$ SCFT with $SO(2)_R$ symmetry for $a=1,5,10$ (red, green, blue).}\label{fig5}
 \end{figure}

We then move on and briefly look at the $SO(2)_R$ symmetry. There are three singlet scalars from the $SO(3,3)/SO(3)\times SO(3)$ coset. These scalars will be denoted by $\Phi_1$, $\Phi_2$ and $\Phi_3$ corresponding to the non-compact generators $Y_{31}$, $Y_{32}$ and $Y_{33}$, respectively.
\\
\indent In this case, the gauge field corresponding the $SO(2)_R$ generator is given by
\begin{equation}
A^3=a\cos\theta d\phi\, .
\end{equation}
By using the same procedure, we find that, in order to have a fixed point, all of the $\Phi_i$'s must vanish, and only $AdS_5\times H^2$ solutions exist. The solution again preserves eight supercharges corresponding to $N=1$ superconformal symmetry in four dimensions. The fixed point solution is given by
\begin{equation}
\sigma=\frac{2}{5}\ln \frac{4}{3},\qquad G=\frac{1}{5}\ln \frac{4}{3}-\frac{1}{2}\ln \frac{g_1}{3a},\qquad
F=\frac{16h}{9^\frac{2}{5}}r
\end{equation}
\indent There exist RG flows from the $SO(4)$ $N=(1,0)$ SCFT to these four-dimensional SCFTs. The BPS equations describing theses flows are given by
\begin{eqnarray}
\sigma'&=&\frac{2}{5}e^{-\frac{\sigma}{2}}\left(ae^{\sigma-2G}-g_1-16he^{\frac{5\sigma}{2}}\right),\\
G'&=&\frac{1}{5}e^{-\frac{\sigma}{2}}\left(4he^{\frac{5\sigma}{2}}-g_1-4ae^{\sigma-2G}\right),\\
F'&=&\frac{1}{5}e^{-\frac{\sigma}{2}}\left(4he^{\frac{5\sigma}{2}}-g_1+ae^{\sigma-2G}\right).
\end{eqnarray}
Examples of the solutions with some values of the parameter $a$ are shown in Figure \ref{fig5}. This critical point is also a solution of pure $N=2$ gauged supergravity studied in \cite{MN_nogo}.

\section{Flows to $N=1$ SCFTs in three dimensions}\label{AdS4}
In this section, we look for $AdS_4$ vacua of the form $AdS_4\times S^3$ or $AdS_4\times H^3$ with $S^3$ and $H^3$ being a three-sphere and a three-dimensional hyperbolic space, respectively. These solutions will correspond to some SCFTs in three dimensions. In order to identify these $AdS_4$ vacua with the IR fixed points of the six-dimensional SCFTs corresponding to both of the $AdS_7$ vacua given in \eqref{SO4_AdS7} and \eqref{SO3_AdS7}, we consider the scalars which are singlets under $SO(3)_{\textrm{diag}}$ subgroup of the full $SO(4)$ gauge group. The relevant scalar from the $SO(3,3)/SO(3)\times SO(3)$ coset is the one corresponding to the generator \eqref{SO3_singlet_scalar} with the coset representative given in \eqref{SO3_diag_L}.
\\
\indent
In the $S^3$ case, we will take the metric ansatz to be
\begin{equation}
ds^2_7=e^{2F}dx^2_{1,2}+dr^2+e^{2G}\left[d\psi^2+\sin^2\psi (d\theta^2+\sin^2\theta d\phi^2)\right].
\end{equation}
From the above metric, we find the spin connections
\begin{eqnarray}
\omega^{\hat{\mu}}_{\phantom{s}\hat{r}}&=&F'e^{\hat{\mu}},\qquad \omega^{\hat{\psi}}_{\phantom{s}\hat{r}}=G'e^{\hat{\psi}},\qquad
\omega^{\hat{\theta}}_{\phantom{s}\hat{r}}=G'e^{\hat{\theta}},\nonumber \\
\omega^{\hat{\phi}}_{\phantom{s}\hat{r}}&=&G'e^{\hat{\phi}},\qquad \omega^{\hat{\phi}}_{\phantom{s}\hat{\theta}}=e^{-G}\frac{\cot\theta}{\sin \psi}e^{\hat{\phi}},\nonumber \\
\omega^{\hat{\phi}}_{\phantom{s}\hat{\psi}}&=&e^{-G}\cot\psi e^{\hat{\phi}},\qquad
\omega^{\hat{\theta}}_{\phantom{s}\hat{\psi}}=e^{-G}\cot\psi e^{\hat{\theta}}
\end{eqnarray}
which accordingly suggest to turn on the following $SO(3)_{\textrm{diag}}$ gauge fields
\begin{eqnarray}
A^1&=&\frac{g_2}{g_1}A^4=a\cos\psi d\theta,\nonumber \\
A^2&=&\frac{g_2}{g_1}A^5=a\cos\theta d\phi,\nonumber \\
A^3&=&\frac{g_2}{g_1}A^6=a\cos\psi\sin \theta d\phi\, .
\end{eqnarray}
Note that at the beginning, the parameter $a$ of each gauge field needs not be equal. However, the twist
condition
\begin{equation}
ag_1=1\label{twist_S3}
\end{equation}
requires that all of the parameters in front of $A^i$ must be equal. The corresponding field strengths are, after using \eqref{twist_S3},
\begin{eqnarray}
F^1&=&-ae^{-2G}e^{\hat{\psi}}\wedge e^{\hat{\theta}},\nonumber\\
F^2&=&-ae^{-2G}e^{\hat{\theta}}\wedge e^{\hat{\phi}},\nonumber\\
F^3&=&-ae^{-2G}e^{\hat{\psi}}\wedge e^{\hat{\phi}}\, .
\end{eqnarray}
\indent To set up the BPS equations, we impose the projection conditions
\begin{equation}
\gamma_r\epsilon =\epsilon,\qquad i\sigma^1\gamma_{\hat{\theta}\hat{\psi}}\epsilon=\epsilon,\qquad
i\sigma^2\gamma_{\hat{\phi}\hat{\theta}}\epsilon =\epsilon,\qquad i\sigma^3\gamma_{\hat{\phi}\hat{\psi}}\epsilon=\epsilon\, .\label{projector_S3}
\end{equation}
\indent
For the $H^3$ case, we take the metric to be
\begin{equation}
ds^2_7=e^{2F}dx^2_{1,2}+dr^2+\frac{e^{2G}}{y^2}(dx^2+dy^2+dz^2)\label{AdS4H3_metric}
\end{equation}
with the spin connections given by
\begin{eqnarray}
\omega^{\hat{z}}_{\phantom{s}\hat{r}}&=&G'e^{\hat{z}},\qquad \omega^{\hat{y}}_{\phantom{s}\hat{r}}=G'e^{\hat{y}},\qquad \omega^{\hat{x}}_{\phantom{s}\hat{r}}=G'e^{\hat{x}},\nonumber \\
\omega^{\hat{x}}_{\phantom{s}\hat{y}}&=&-e^{-G}e^{\hat{x}},\qquad \omega^{\hat{z}}_{\phantom{s}\hat{y}}=-e^{-G}e^{\hat{z}},\qquad \omega^{\hat{\mu}}_{\phantom{s}\hat{r}}=F'e^{\hat{\mu}}\, .
\end{eqnarray}
We then turn on the following gauge fields, to cancel the above spin connections on $H^3$,
\begin{equation}
A^1=\frac{a}{y}dx,\qquad A^2=0,\qquad A^3=\frac{a}{y}dz
\end{equation}
with $A^{i+3}=\frac{g_1}{g_2}A^i$, $i=1,2,3$. These gauge fields then become $SO(3)_{\textrm{diag}}$ gauge fields. \\
\indent
We will also impose the projection conditions
\begin{equation}
\gamma_r\epsilon =\epsilon,\qquad i\sigma^1\gamma_{\hat{x}\hat{y}}\epsilon=-\epsilon,\qquad
i\sigma^2\gamma_{\hat{x}\hat{z}}\epsilon =-\epsilon,\qquad i\sigma^3\gamma_{\hat{z}\hat{y}}\epsilon=-\epsilon\, .\label{projector_H3}
\end{equation}
The twist condition is still given by \eqref{twist_S3}.
\\
\indent In both cases, the last projector in \eqref{projector_S3}
and \eqref{projector_H3} is not independent from the second and the
third ones, so the fixed point solution will preserve four
supercharges corresponding to $N=1$ superconformal symmetry in three
dimensions.
\\
\indent With all of the above conditions, we find the following BPS
equations, for the $H^3$ case,
\begin{eqnarray}
\phi'&=&-\frac{1}{8g_2}e^{-\frac{\sigma}{2}-3\phi-2G}\left[e^{2G}(e^{4\phi}-1)g_2-4ae^{\sigma+2\phi}\right]
\left[g_1-g_2+(g_1+g_2)e^{2\phi}\right],\quad\\
\sigma'&=&-\frac{1}{20}e^{-\frac{\sigma}{2}-3\phi-2G}\left[\frac{12a}{g_2}e^{\sigma+2\phi}
\left[(e^{2\phi}-1)g_1
+(1+e^{2\phi})g_2\right]\right.\nonumber \\
& &\left.+e^{2G}g_2\left[ g_2(e^{2\phi}-1)^3+g_1(e^{2\phi}+1)^3
+128he^{\frac{5\sigma}{2}+3\phi}\right]\right],\\
G'&=&\frac{1}{40}e^{-\frac{\sigma}{2}-3\phi-2G}\left[\frac{28a}{g_2}e^{\sigma+2\phi}\left[(e^{2\phi}-1)g_1
+(1+e^{2\phi})g_2\right]\right.\nonumber \\
& &\left.-e^{2G}g_2\left[ g_2(e^{2\phi}-1)^3+g_1(e^{2\phi}+1)^3
-32he^{\frac{5\sigma}{2}+3\phi}\right]\right],\\
F'&=&-\frac{1}{40}e^{-\frac{\sigma}{2}-3\phi-2G}\left[\frac{12a}{g_2}e^{\sigma+2\phi}\left[(e^{2\phi}-1)g_1
+(1+e^{2\phi})g_2\right]\right.\nonumber \\
& &\left.+e^{2G}g_2\left[ g_2(e^{2\phi}-1)^3+g_1(e^{2\phi}+1)^3
-32he^{\frac{5\sigma}{2}+3\phi}\right]\right].
\end{eqnarray}
The corresponding equations for the $S^3$ case are similar with $a$
replaced by $-a$.
\\
\indent We now look for a fixed point solution at which
$G'=\phi'=\sigma'=0$ and $F'=\textrm{constant}$. For $g_2=g_1$, only
$AdS_4\times H^3$ solutions exist and are given by
\begin{eqnarray}
\phi&=&\frac{1}{4}\ln 2,\qquad \sigma =\frac{3}{10}\ln 2,\nonumber
\\
G&=&\frac{1}{10}\ln \left[\frac{64a^5}{g_1^3h^2}\right],\qquad
L_{AdS_5}=\frac{1}{2^{\frac{13}{5}}h}\, .\label{AdS4S3_7D}
\end{eqnarray}
This solution can be uplifted to eleven dimensions using the ansatz
of \cite{SO4_7Dfrom11D}.
\\
\indent When $g_2\neq g_1$, we also find $AdS_4\times H^3$ solutions
\begin{eqnarray}
\phi&=&\frac{1}{2}\ln \left[\frac{g_2-g_1}{g_2+g_1}\right],\qquad \sigma=\frac{1}{5}\ln \left[\frac{g_1^2g_2^2}{100h^2(g_2^2-g_1^2)}\right],\nonumber \\
G&=&\frac{1}{2}\ln \left[\frac{5a(g_2^2-g_1^2)}{g_1g_2^2}\right]+\frac{1}{5}\ln \left[\frac{-g_1g_2}{10h\sqrt{g_2^2-g_1^2}}\right],\nonumber \\
L_{AdS_4}&=&\frac{1}{2^{\frac{6}{5}}h}\left[\frac{25h^2(g_2^2-g_1^2)}{g_1^2g_2^2}\right]^{\frac{2}{5}}.
\end{eqnarray}
This solution can be connected to both $AdS_7$ critical points in \eqref{SO4_AdS7} and \eqref{SO3_AdS7} by some RG flows.
\\
\indent In this $g_2\neq g_1$ case, there can be both $AdS_4\times S^3$ and $AdS_4\times
H^3$ solutions. The solution however takes a more complicated form depending on the values of $g_1$ and $g_2$.
The $AdS_4\times H^3$ and $AdS_4\times S^3$ solutions are given respectively by
\begin{eqnarray}
G &=&\frac{1}{2}\ln \left[\frac{4ae^{\sigma+2\phi_0}}{g_2(e^{4\phi_0}-1)}\right],\\
\sigma&=&\frac{2}{5}\ln\left[\frac{e^{-3\phi_0}\left[g_2(1-e^{6\phi_0})-g_1(e^{6\phi_0}+1)\right]}{32h}\right]
\end{eqnarray}
and
\begin{eqnarray}
G &=&\frac{1}{2}\ln \left[\frac{4ae^{\sigma+2\phi_0}}{g_2(1-e^{4\phi_0})}\right],\\
\sigma&=&\frac{2}{5}\ln\left[\frac{e^{-3\phi_0}\left[g_2(1-e^{6\phi_0})-g_1(e^{6\phi_0}+1)\right]}{32h}\right].
\end{eqnarray}
\indent In both cases, the scalar $\phi_0$ is a solution to the equation
\begin{equation}
g_1(1-2e^{2\phi_0}-2e^{4\phi_0}+e^{6\phi_0})-g_2(1+2e^{2\phi_0}-2e^{4\phi_0}-e^{6\phi_0})=0\,
.\label{phi0}
\end{equation}
\indent The explicit form of $\phi_0$ can be obtained but will not be given
here due to its complexity. There are many possible solutions for
$\phi_0$ depending on the values of $g_1$, $g_2$ and $a$. An example
of $AdS_4\times S^3$ solutions is, for $g_2=\frac{1}{2}g_1$, given
by
\begin{equation}
\phi=-0.9158,\qquad \sigma=0.5493,\qquad G = 0.4116+\frac{1}{2}\ln
\left[\frac{a}{g_1}\right]\, .
\end{equation}
One of the $AdS_4\times H^3$ solutions is, for $g_2=\frac{1}{2}g_1$, given by
\begin{equation}
\phi=0.2706,\qquad \sigma=0.2351,\qquad G = 1.0936+\frac{1}{2}\ln
\left[\frac{a}{g_1}\right]\, .
\end{equation}
\indent Numerical solutions for RG flows from the UV $N=(1,0)$ SCFTs
in six dimensions to these three-dimensional $N=1$ SCFTs can be
found in the same way as those given in the previous section. And,
with suitable boundary conditions, the flow from $SO(4)$ $AdS_7$
point to the $SO(3)$ $AdS_7$ point and then to $AdS_4\times S^3$ or
$AdS_4\times H^3$ in the case of $g_2\neq g_1$ should be similarly
obtained. We will however not give these solutions here.
\section{Uplifting the solutions to eleven dimensions}\label{uplift11D}
In this section, we will uplift some of the $AdS_5$ and $AdS_4$ solutions found in the previous sections to eleven dimensions using a reduction ansatz given in \cite{SO4_7Dfrom11D}. Only solutions with equal $SU(2)$ gauge couplings, $g_2=g_1$, can be uplifted by this ansatz. Therefore, we will consider only this case in the remaining of this section.
\\
\indent The reduction ansatz given in \cite{SO4_7Dfrom11D} is naturally written in terms of $SL(4,\mathbb{R})/SO(4)$ scalar manifold rather than the $SO(3,3)/SO(3)\times SO(3)$ we have considered throughout the previous sections. It is then useful to change the parametrization of scalars from the $SO(3,3)/SO(3)\times SO(3)$ to $SL(4,\mathbb{R})/SO(4)$ cosets. For convenience, we will repeat the supersymmetry transformations of fermions with the three-form field and fermions vanishing
\begin{eqnarray}
\delta \psi_\mu &=&D_\mu \epsilon -\frac{1}{20}gX\tilde{T}\gamma_\mu
\epsilon-\frac{1}{20}X^{-4}\gamma_\mu\epsilon\nonumber \\
&
&+\frac{1}{40\sqrt{2}}X^{-1}\left(\gamma_\mu^{\phantom{s}\nu\rho}-8\delta^\nu_\mu
\gamma^\rho\right)\Gamma_{RS}F^{RS}_{\nu\rho}\epsilon,\label{delta1}\\
\delta\chi &=&-X^{-1}\gamma^\mu\pd_\mu X
\epsilon-\frac{2}{5}gX^{-4}\epsilon+\frac{1}{10}gX\tilde{T}-\frac{1}{20\sqrt{2}}X^{-1}
\gamma^{\mu\nu}\Gamma_{RS}F^{RS}_{\mu\nu}\epsilon,\quad \label{delta2}\\
\delta\hat{\lambda}_R&=&-\frac{1}{2}\gamma^\mu\Gamma^S P_{\mu
RS}\epsilon-\frac{1}{8}gX\tilde{T}\Gamma_R \epsilon
+\frac{1}{2}gX\tilde{T}_{RS}\Gamma^S\epsilon \nonumber \\
&
&-\frac{1}{8\sqrt{2}}X^{-1}\gamma^{\mu\nu}\Gamma_S\left(F^{RS}_{\mu\nu}
+\frac{1}{2}\epsilon_{RSTU}F^{TU}_{\mu\nu}\right)\epsilon
\label{delta3}
\end{eqnarray}
where
\begin{eqnarray}
P_{RS}&=&(\mc{V}^{-1})^\alpha_{(R}\left(\delta^\beta_\alpha
d+gA_{(1)\alpha} ^{\phantom{sdss}\beta}\right)\mc{V}_\beta
^{\phantom{s}T}\delta_{S)T},\nonumber \\
Q_{RS}&=&(\mc{V}^{-1})^\alpha_{[R}\left(\delta^\beta_\alpha
d+gA_{(1)\alpha} ^{\phantom{sdss}\beta}\right)\mc{V}_\beta
^{\phantom{s}T}\delta_{S]T},\nonumber \\
D\epsilon &=& d\epsilon
+\frac{1}{4}\omega_{ab}\gamma^{ab}+\frac{1}{4}Q_{RS}\Gamma^{RS}\nonumber
\\
\tilde{T}_{RS}&=&(\mc{V}^{-1})_R^{\phantom{s}\alpha}(\mc{V}^{-1})_S^{\phantom{s}\beta}\delta_{\alpha\beta},
\qquad \tilde{T}=\tilde{T}_{RS}\delta^{RS}\, .
\end{eqnarray}
In the above equations, $\mc{V}^R_{\phantom{s}\alpha}$ denotes the
$SL(4,\mathbb{R})/SO(4)$ coset representative.
\\
\indent For the explicit form of the eleven-dimensional metric and
the four-form field including the notations used in the above equations, we refer the reader to
\cite{SO4_7Dfrom11D}. We now consider the $AdS_5$ and $AdS_4$
solutions separately.

\subsection{Uplifting the $AdS_5$ solutions}
For $AdS_5$ solutions, the seven-dimensional metric is given by \eqref{AdS5S2_metric} and \eqref{metric_AdS5_H2}.
We will restrict ourselves to $AdS_5$ fixed points with $SO(2)\times SO(2)$ symmetry. The non-zero gauge fields are $A^{\alpha\beta}=(A^{12},A^{34})$ whose explicit form is given by
\begin{equation}
A^{12}=a\cos\theta d\phi\qquad\textrm{and}\qquad A^{34}=b\cos\theta d\phi\, .
\end{equation}
The $U(1)\times U(1)$ singlet scalar from $SL(4,\mathbb{R})/SO(4)$ coset is parametrized by the
coset representative
\begin{equation}
\mc{V}^R_{\phantom{s}\alpha}=
\textrm{diag}(e^{\frac{\Phi}{2}},e^{\frac{\Phi}{2}},e^{-\frac{\Phi}{2}},e^{-\frac{\Phi}{2}})
\end{equation}
from which the
$\tilde{T}_{RS}=\textrm{diag}(e^{-\Phi},e^{-\Phi},e^{\Phi},e^{\Phi})$
follows. Note that the parameter $a$ and $b$ here are different from
those in section \ref{AdS5} since the gauge fields $A^i$ and $A^r$
correspond respectively to the anti-self-dual and self-dual parts of
the $SO(4)$ gauge fields $A^{\alpha\beta}$.
\\
\indent Using the above supersymmetry transformations and imposing
the projection conditions $\gamma_{\hat{r}}\epsilon=\epsilon$ and
$\gamma^{\hat{\theta}\hat{\phi}}\Gamma_{12}\epsilon=\epsilon$, we
obtain the BPS equations
\begin{eqnarray}
X^{-1}X'-\frac{2}{5}g
X^{-4}+\frac{1}{5}gX(e^{\Phi}+e^{-\Phi})+\frac{1}{5\sqrt{2}}X^{-1}e^{-2G}(ae^\Phi-be^{-\Phi})&=&0,\label{SO2SO2_eq1}\\
-\Phi'-gX(e^\Phi-e^{-\Phi})+\frac{1}{\sqrt{2}}X^{-1}e^{-2G}(ae^\Phi+be^{-\Phi})&=&0,\label{SO2SO2_eq2}\\
F'-\frac{1}{5}gX(e^\Phi+e^{-\Phi})-\frac{1}{10}gX^{-4}-\frac{1}{10\sqrt{2}}X^{-1}
e^{-2G}(ae^\Phi-be^{-\Phi})&=&0,\label{SO2SO2_eq3}\\
G'-\frac{1}{5}gX(e^\Phi+e^{-\Phi})-\frac{1}{10}gX^{-4}+\frac{4}{5\sqrt{2}}X^{-1}
e^{-2G}(ae^\Phi-be^{-\Phi})&=&0\, .\label{SO2SO2_eq4}
\end{eqnarray}
In the above equations, we have used $\Gamma_{34}\epsilon=-\Gamma_{12}\epsilon$ which follows from
the condition $\Gamma_{1234}\epsilon =\epsilon$. The latter is part of the truncation from the maximal $SO(5)$ gauged supergravity to the half-maximal $SO(4)$ gauged supergravity studied in \cite{SO4_7Dfrom11D}. We have also used the twist condition given by
\begin{equation}
g(a-b)+1=0\, .\label{AdS5_twist_SL4}
\end{equation}
which comes from the requirement that the gauge connection cancels the spin connection. Note that this condition differs from \eqref{S2_twist} since the gauge fields are different. In condition \eqref{S2_twist}, the $SU(2)_R$ gauge fields are given by the $A^I$ with $I=1,2,3$, and the $SO(2)_R\subset SU(2)_R$ gauge field has been chosen to be $A^3$. On the other hand, the condition \eqref{AdS5_twist_SL4} involves $A^{12}-A^{34}$ corresponding to the $SO(2)_R$ subgroup of the $SU(2)_R$ R-Symmetry for which the corresponding gauge fields are identified with the anti-self-dual part of the $SO(4)$ gauge fields $A^{\alpha\beta}$ in the convention of \cite{SO4_7Dfrom11D}.
\\
\indent For large $r$, the solution should approach $X=1$, $\Phi=0$ and $F\sim G\sim r$ giving
$AdS_7$ background with $SO(4)$ symmetry. This corresponds to the UV
$N=(1,0)$ SCFT in six dimensions. In the IR with the boundary
condition $F\sim r$ and $G, \Phi, \sigma\sim \textrm{constant}$, there is a class of
solutions given by
\begin{eqnarray}
\Phi &=&\frac{1}{2}\ln
\left[\frac{a+b\pm\sqrt{a^2+ab+b^2}}{a}\right],\nonumber \\
G&=&\frac{1}{2}\ln
\left[\frac{a\left(a+2b\pm \sqrt{a^2+ab+b^2}\right)}{\sqrt{2}gX^2\left(b\pm\sqrt{a^2+ab+b^2}\right)}
\right],\nonumber \\
X^{10}&=&\frac{a\left(a+2b\pm\sqrt{a^2+ab+b^2}\right)^2}{4(a+b)^2\left(a+b
\pm\sqrt{a^2+ab+b^2}\right)},\nonumber \\
L_{AdS_5}&=&\frac{a2^{\frac{1}{5}}}{g}\left[\frac{a+2b\pm \sqrt{a^2+ab+b^2}}
{(a+b)^2\left(a+b\pm \sqrt{a^2+ab+b^2}\right)}\right]^{\frac{2}{5}}\, .\label{AdS5_SL4}
\end{eqnarray}
This gives $AdS_5\times S^2$ background preserving $U(1)\times U(1)$
symmetry and eight supercharges since only the projector $\gamma^{\hat{\theta}\hat{\phi}}\Gamma_{12}\epsilon=\epsilon$ is needed at the fixed point. Therefore, this solution corresponds to $N=1$ SCFT in four dimensions. This solution is the same as in \cite{Cucu1} with the identification
$(m_1,m_2)\rightarrow (-b,a)$ up to some field redefinitions. So, we conclude that the $AdS_5\times \Sigma_2$ solutions found in \cite{Cucu1} is a solution of the $N=2$ $SO(4)$ gauged supergravity.
\\
\indent For the $H^2$ case, the above analysis can be repeated in a similar manner. The resulting BPS equations are, as expected, given by \eqref{SO2SO2_eq1}, \eqref{SO2SO2_eq2}, \eqref{SO2SO2_eq3} and \eqref{SO2SO2_eq4} with $(a,b)$ replaced by $(-a,-b)$. It can also be verified that for both $AdS_5\times S^2$ and $AdS_5\times H^2$ solutions given in \eqref{AdS5_SL4}, solutions with the positive sign are valid for $g>0$ and $a>0$ while solutions with the negative sign are valid for $g<0$ and $a<0$.
\\
\indent
It should also be noted that we can truncate the above BPS equations to those of $SO(2)_R$ symmetry, generated by the anti-selfdual gauge field $A^{12}-A^{34}$, by setting $b=-a$. Since the twist condition in this case becomes $2ga=-1$ which implies that $ga<0$, only the $AdS_5\times H^2$ exists. This precisely agrees with the result of section \ref{SO2R_flow}. The corresponding solution is given by
\begin{equation}
X=\left(\frac{3}{4}\right)^{\frac{1}{5}},\qquad G=-\frac{1}{2}\ln \left[-\frac{g}{2^{\frac{3}{10}}3^{\frac{3}{5}}a}\right],\qquad L_{AdS_5}=\frac{3^{\frac{4}{5}}}{2^{\frac{3}{5}}g}\, .
\end{equation}
\indent The $AdS_5\times H^2$ with $SO(2)_{\textrm{diag}}$ symmetry
found in section \ref{SO2_DAdS51} for $g_2=g_1$ can also be uplifted
using the formulae given here by truncating the $SO(2)\times SO(2)$
symmetry to $SO(2)_{\textrm{diag}}$ as remarked previously in
section \ref{SO2_DAdS51}. The $SO(2)_{\textrm{diag}}$ corresponds to
the gauge field $A^{12}$ since the $A^3$ and $A^6$, in section
\ref{SO2_DAdS5}, are related to the anti-self-dual,
$\frac{1}{2}(A^{12}-A^{34})$, and self-dual,
$\frac{1}{2}(A^{12}+A^{34})$, fields, respectively. So, the
$SO(2)_{\textrm{diag}}$ gauge field is given by $A^{12}$. As in section \ref{SO2_DAdS5}, only solutions with the upper
sign in the solution \eqref{AdS5_SL4} and $AdS_5\times H^2$ are
possible. The result is given by
\begin{equation}
\Phi=\frac{1}{2}\ln 2,\qquad X^{10}=\frac{1}{8},\qquad
G=\frac{1}{2}\ln \left[-\frac{a2^{\frac{11}{10}}}{g}\right].
\end{equation}
This is consistent with the twist condition \eqref{AdS5_twist_SL4}
which, for $b=0$, becomes $ga=-1$.
\\
\indent We now move to the uplift of these $AdS_5$ solutions. Both $AdS_5\times S^2$ and
$AdS_5\times H^2$ solutions can be uplifted in a similar way. For
definiteness, we will only give the uplifted $AdS_5\times S^2$
solution. Using the reduction ansatz given in \cite{SO4_7Dfrom11D}, we
find the eleven-dimensional metric
\begin{eqnarray}
ds^2_{11}&=&\Delta^{\frac{1}{3}}\left[e^{\frac{2r}{L_{AdS_5}}}dx^2_{1,3}+dr^2+e^{2G_0}(d\theta^2+\sin^2\theta
d\phi^2)\right]\nonumber \\
&
&+\frac{2}{g}\Delta^{-\frac{2}{3}}X_0^3\left[X_0\cos^2\xi+X_0^{-4}\sin^2\xi
\left(e^{-\Phi_0}\sin^2\psi
+e^{\Phi_0}\cos^2\psi\right)\right]d\xi^2\nonumber \\
& &+\frac{1}{2g^2}\Delta^{-\frac{2}{3}}X_0^{-1}\cos^2\xi
\left[e^{-\Phi_0}\left[\cos^2\psi d\phi^2+\sin^2\psi
(d\alpha-ag\cos\theta d\phi)^2\right]\right.\nonumber
\\
& &+\left.e^{\Phi_0}\left[\cos^2\psi d\phi^2+\sin^2\psi
(d\beta-bg\cos\theta d\phi)^2\right]\right]\nonumber \\
& &-\frac{1}{2g^2}\Delta^{-\frac{2}{3}}X_0^{-1}\sin \xi \sin
(2\psi)\left(e^{-\Phi_0}-e^{\Phi_0}\right)d\xi d\psi
\end{eqnarray}
where we have used the coordinates $\mu^\alpha$, satisfying $\mu^\alpha\mu^\alpha=1$, as follow
\begin{eqnarray}
\mu^1&=&\sin \psi \cos \alpha,\qquad \mu^2=\sin \psi \sin
\alpha,\nonumber \\
\mu^3&=&\cos\psi \cos \beta,\qquad \mu^4=\cos \psi\sin\beta\, .
\end{eqnarray}
The quantities $X_0$, $\Phi_0$ and $G_0$ are the values of the
corresponding fields at the fixed point \eqref{AdS5_SL4}. The
quantity $\Delta$ is defined by
\begin{equation}
\Delta
=X^{-4}\sin^2\xi+X\tilde{T}_{\alpha}\mu^\alpha\mu^\beta\cos^2\xi
\end{equation}
which, in the present case, gives
\begin{equation}
\Delta=X_0\cos^2\xi \left(e^{-\Phi_0}\sin^2\psi
+e^{\Phi_0}\cos^2\psi \right)+X_0^{-4}\sin^2\xi\, .
\end{equation}
\indent The 4-form field, at the fixed point, is given by
\begin{eqnarray}
\hat{F}_{(4)}&=&\frac{1}{g^3}U\Delta^{-2}\cos^3\xi d\xi \wedge
\epsilon_{(3)}+\frac{a}{g^2}\cos\theta\cos\xi
\left[\sin\xi \cos\xi \sin\psi \cos\psi X_0^{-4}d\psi \right.\nonumber \\
& &\left. \cos^2\psi \left(X_0^{-4}\sin^2\xi +e^{\Phi_0}X_0^2\cos^2\xi \right)d\xi\right]\wedge
d\beta\wedge d\theta\wedge d\phi\nonumber \\
& &-\frac{b}{g^2}\sin\theta\cos\xi\left[\sin\xi\cos\xi\sin\psi\cos\psi X_0^{-4}d\psi \right.\nonumber \\
& & \left.-\left(X_0^{-4}\sin^2\xi +X_0^2\cos^2\xi
e^{-\Phi_0}\right)\sin^2\psi d\xi\right]\wedge d\alpha\wedge
d\theta\wedge d\phi
\end{eqnarray}
where
\begin{eqnarray}
U&=&\sin^2\xi \left[X_0^{-8}-2X_0^{-3}\left(e^{\Phi_0}+e^{-\Phi_0}\right)\right]\nonumber \\
& &-\cos^2\xi\left[2X_0^2+X_0^{-3}\left(e^{-\Phi_0}\sin^2\psi
+e^{\Phi_0}\cos^2\psi\right)\right].
\end{eqnarray}
The uplifted solutions for some particular values of $a$ and $b$ have already been given in \cite{Cucu2}.

\subsection{Uplifting the $AdS_4$ solutions}
We now consider the embedding of the $AdS_4\times H^3$ solution
given in \eqref{AdS4S3_7D} in eleven dimensions. The
$SL(4,\mathbb{R})/SO(4)$ coset representative, invariant under
$SO(3)_{\textrm{diag}}$, is given by
\begin{equation}
\mc{V}^R_{\phantom{s}\alpha}=(\delta_{ab}e^{\frac{\phi}{2}},e^{-\frac{3\phi}{2}})
\end{equation}
which gives
$\tilde{T}_{RS}=(\delta_{ab}e^{-\phi},e^{3\phi})$. We have
split the $\alpha$ index as follow $\alpha=(a,4)$, $a=1,2,3$.
\\
\indent To set up the associated BPS equations, we use the
seven-dimensional metric \eqref{AdS4H3_metric} and the following
gauge fields
\begin{equation}
A^{12}=\frac{a}{y}dz,\qquad A^{31}=0, \qquad A^{23}=\frac{a}{y}dx\,
.
\end{equation}
The twist condition is given by $ga=1$. We will also impose the
projection conditions
\begin{equation}
\Gamma_{23}\gamma_{\hat{x}\hat{y}}\epsilon=-\epsilon,\qquad
\Gamma_{13}\gamma_{\hat{z}\hat{x}}\epsilon=-\epsilon,\qquad
\Gamma_{12}\gamma_{\hat{z}\hat{y}}\epsilon=-\epsilon,\qquad
\Gamma_{\hat{r}}\epsilon=\epsilon\, .
\end{equation}
\indent With all of the above conditions, we obtain the following BPS
equations
\begin{eqnarray}
-\phi'+\frac{1}{2}gX(e^{-\phi}-e^{3\phi})+\sqrt{2}aX^{-1}e^{\phi-2G}&=&0,\\
-X^{-1}X'-\frac{2}{5}gX^{-4}+\frac{1}{10}gX(3e^{-\phi}+e^{3\phi})+\frac{3}{5\sqrt{2}}aX^{-1}
e^{\phi-2G}&=&0,\\
G'-\frac{1}{10}gX(3e^{-\phi}+e^{3\phi})-\frac{1}{10}gX^{-4}+\frac{7}{5\sqrt{2}}aX^{-1}e^{\phi-2G}&=&0,\\
F'-\frac{1}{10}gX(3e^{-\phi}+e^{3\phi})-\frac{1}{10}gX^{-4}-\frac{3}{5\sqrt{2}}aX^{-1}e^{\phi-2G}&=&0\,
.
\end{eqnarray}
These equations admit a fixed point solution
\begin{eqnarray}
\phi_0&=&\frac{1}{4}\ln \frac{11}{3},\qquad
X_0^{20}=\frac{11(3^3)}{2^{12}},\nonumber \\
G_0&=&\frac{1}{10}\ln
\left[\frac{3(11^2)}{2\sqrt{2}}\right]-\frac{1}{2}\ln\left[\frac{g}{a}\right],\qquad
L_{AdS_4}=\frac{1}{g}\left(\frac{11(3^3)}{2^7}\right)^{\frac{1}{5}}\,
.
\end{eqnarray}
\indent The parametrization of the $\mu^\alpha$ coordinates can be
chosen to be
\begin{equation}
\mu^\alpha=(\cos\Psi\hat{\mu}^a,\sin\Psi)
\end{equation}
with $\hat{\mu}^a$ satisfying $\hat{\mu}^a\hat{\mu}^a=1$. The
$SO(3)_{\textrm{diag}}$ symmetry corresponds to the gauge fields
$A^{ab}$. In the following, we accordingly set $A^{4a}=0$ for
$a=1,2,3$ and find that
\begin{equation}
D\mu^a=\cos\Psi D\hat{\mu}^a-\sin\Psi \hat{\mu}^ad\Psi,\qquad
D\mu^4=\cos\Psi d\Psi
\end{equation}
where
\begin{equation}
D\hat{\mu}^a=d\hat{\mu}^a+gA^{ab}\hat{\mu}^b\, .
\end{equation}
With all these results, the eleven-dimensional metric is given by
\begin{eqnarray}
ds^2_{11}&=&\Delta^{\frac{1}{3}}\left[e^{\frac{r}{L_{AdS_4}}}dx^2_{1,2}+dr^2+\frac{e^{2G_0}}{y^2}
\left[dx^2+dy^2+dz^2\right]\right]\nonumber \\
& &+\frac{2}{g^2}\Delta^{-\frac{2}{3}}X_0^3\left[X_0\cos^2\xi +X_0^{-4}\sin^2\xi \left(\cos^2\Psi
e^{\phi_0}+\sin^2\Psi e^{-3\phi_0} \right)\right]d\xi^2\nonumber \\
& &+\frac{1}{2g^2}\Delta^{-\frac{2}{3}}X_0^{-1}\cos^2\xi \left[\cos^2\Psi e^{\phi_0}D\hat{\mu}^aD\hat{\mu}^a
+\left(\sin^2\Psi e^{\phi_0}+\cos^2\Psi e^{-3\phi_0}\right)d\Psi^2\right]\nonumber \\
&
&-\frac{1}{g^2}\Delta^{-\frac{2}{3}}X^{-1}_0\sin\xi\left(e^{-3\phi_0}-e^{\phi_0}\right)\sin\Psi\cos\Psi
d\Psi d\xi\, .
\end{eqnarray}
The $S^2$ coordinates $\hat{\mu}^a$ can be parametrized by
\begin{equation}
\hat{\mu}^1=\sin\beta \cos \alpha,\qquad \hat{\mu}^2=\sin\beta \sin \alpha,\qquad \hat{\mu}^3=\cos \beta\, .
\end{equation}
The warped factor $\Delta$ is given by
\begin{equation}
\Delta = X^2_0e^{-\phi_0}\cos^2\xi\cos^2\Psi+X_0^{-4}\sin^2\xi
+X_0e^{3\phi_0}\sin^2\Psi \cos^2\xi\, .
\end{equation}
\indent The four-form field on the $AdS_4\times H^3$ background can be written as
\begin{eqnarray}
\hat{F}_{(4)}&=&\frac{1}{g^3}U\cos^3\xi \cos^2\Psi d\xi\wedge d\Psi \wedge \epsilon_{(2)}\nonumber \\
& &+\frac{1}{2g^2}\cos\xi \epsilon_{abc}\left[\hat{\mu}^c\left[X^{-4}_0\sin^2\xi (\sin^2\Psi-\cos^2\Psi)\right.\right.
\nonumber \\
& &
\left.+X^2_0(e^{3\phi_0}\sin^2\Psi-e^{-\phi_0}\cos^2\Psi)\right]d\xi
\wedge F^{ab}\wedge d\Psi
\nonumber \\
& &-\left[(X^{-4}_0\sin^2\xi +X^2_0\cos^2\xi e^{3\phi_0})\sin\Psi \cos\Psi d\xi\right.
\nonumber \\
& &
\left.\left.+X^{-4}_0\cos \xi \sin\xi \cos^2\Psi d\Psi\right]\wedge F^{ab}\wedge D\hat{\mu}^c  \right]
\end{eqnarray}
where
\begin{eqnarray}
\epsilon_{(2)}&=&\frac{1}{2}\epsilon_{abc}\hat{\mu}^aD\hat{\mu}^b\wedge D\hat{\mu}^c,\nonumber \\
U&=&\cos^2\xi
\left[X_0^2\left[e^{6\phi_0}\sin^2\Psi-e^{-2\phi_0}\cos^2\Psi-e^{2\phi_0}(2\sin^2\Psi+1)\right]\right.
\nonumber \\
& &\left.
-X^{-3}_0(e^{-\phi_0}\cos^2\Psi+e^{3\phi_0}\sin^2\Psi)\right]\nonumber \\
& &+\sin^2\xi X^{-3}_0(X^{-5}_0-3e^{-\phi_0}-e^{3\phi_0}) .
\end{eqnarray}

\section{Conclusions}\label{conclusion}
We have studied $AdS_5\times \Sigma_2$ and $AdS_4\times \Sigma_3$
solutions of $N=2$ gauged supergravity in seven dimensions with
$SO(4)$ gauge group. We have found that there exist both
$AdS_5\times S^2$ and $AdS_5\times H^2$ solutions with the gauge
fields for $SO(2)\times SO(2)$ turned on. With $SO(2)_R$ or
$SO(2)_{\textrm{diag}}$ gauge fields, only $AdS_5\times H^2$
solution is possible. This is consistent with the results given in
\cite{MN_nogo} and \cite{Cucu2}. We recover $AdS_5\times S^2$ and
$AdS_5\times H^2$ solutions studied in \cite{Cucu1} and \cite{Cucu2}
with $SO(2)\times SO(2)$ symmetry. In the case of equal $SU(2)$
gauge couplings, the solutions can be uplifted to eleven dimensions,
and the uplifted solutions have explicitly given.
\\
\indent
We have also considered RG flow solutions
interpolating between supersymmetric $AdS_7$ critical points in the
UV and these $AdS_5$ solutions in the IR. In the case of
$SO(2)_{\textrm{diag}}$ symmetry, there exist flow solutions from
$SO(4)$ $AdS_7$ critical point to $AdS_5$ as well
as flows from $SO(4)$ $AdS_7$ to $SO(3)$ $AdS_7$ and then continue
to $AdS_5$ fixed points similar to the flows from four-dimensional
SCFTs to two-dimensional $N=(2,0)$ SCFTs studied in
\cite{Bobev_4D2D_flow}. Other results of this paper are a number of new $AdS_4\times S^3$ and $AdS_4\times H^3$
solutions for unequal $SU(2)$ gauge couplings. With equal $SU(2)$
couplings, only $AdS_4\times H^3$ geometry is possible, and the
resulting solutions can be uplifted to eleven dimensions.
\\
\indent The results obtained in this paper should be relevant in the
holographic study of $N=(1,0)$ SCFTs in six dimensions. These would
also provide new $AdS_5$ and $AdS_4$ solutions, corresponding to new
SCFTs in four and three dimensions, within the framework of
seven-dimensional gauged supergravity. The embedding of the
solutions in the case of unequal $SU(2)$ gauge couplings (if
possible) would be interesting to explore. It would also be
interesting to compare the $AdS_5$ and $AdS_4$ solutions obtained
here and the solutions found recently in
\cite{Tomasiello_AdS5,Tomasiello_AdS4} in the context of massive
type IIA theory. Finally, it is of particular interest to find an
interpretation of all these solutions in terms of wrapped M5-branes
on $\Sigma_2$ and $\Sigma_3$. Along this line, it would also be
useful to find an implication of the $AdS_4$ solutions in terms of
the M2-brane worldvolume theories.
\acknowledgments
The author would like to thank Carlos Nunez for correspondences
relating to some parts of the results and Eoin O Colgain for
discussions on $AdS_4$ solutions. He is also grateful to Nakwoo Kim
for valuable comments on references. This work is supported by
Chulalongkorn University through Ratchadapisek Sompoch Endowment
Fund under grant RGN-2557-002-02-23. The author is also supported by
The Thailand Research Fund (TRF) under grant TRG5680010.


\end{document}